\documentclass[12pt,a4paper]{article}
\usepackage[text={450pt,650pt},centering]{geometry}
\usepackage{lmodern}
\usepackage{amsmath,amssymb,mathtools,dsfont,booktabs}

\usepackage{hyperref}
\hypersetup{pdftitle={BRST structure of the Higgs-branch induced gauge theory CFT2},
  pdfauthor={Bogdan Stefa\'nski, jr.}}
\usepackage{slashed}

\numberwithin{equation}{section}

\DeclareMathOperator{\tr}{tr}
\newcommand{\alg}[1]{\mathrm{#1}}
\newcommand{\grp}[1]{\mathrm{#1}}
\newcommand{\algSU}{\alg{su}}
\newcommand{\algSO}{\alg{so}}
\newcommand{\algSL}{\alg{sl}}

\newcommand{\sL}{\mbox{\tiny L}}
\newcommand{\sR}{\mbox{\tiny R}}
\newcommand{\superN}{\mathcal{N}}
\newcommand{\CFT}{\text{CFT}}
\newcommand{\Ord}{\mathcal{O}}
\newcommand{\Nc}{N_c}\newcommand{\Nf}{N_f}

\newcommand{\AdS}{\textup{AdS}}
\newcommand{\Sphere}{\mathrm{S}}
\newcommand{\Torus}{\mathrm{T}}

\newcommand{\intq}[1]{\mu^{d-D}\!\int\!\frac{d^{D}#1}{(2\pi)^{D}}\,}
\newcommand{\wreg}{\omega}      
\newcommand{\ereg}{\epsilon}

\usepackage{tikz}
\newcommand{\postarxivchange}[1]{#1}
\usetikzlibrary{calc,decorations.pathmorphing,decorations.markings}
\tikzset{
  scalar/.style={thick,double},
  fermion/.style={thick,double,dashed},
  gluon/.style={thick,double,decorate,decoration={snake,amplitude=.25mm,segment length=1.25mm,post length=0.2mm,pre length=0.2mm}},
  hyp scalar/.style={thick},
  hyp fermion/.style={thick,dashed},
  adj fermion/.style={thick,double,dash pattern=on 0.5mm off 0.5mm},
  aux/.style={thick,double,dotted}
}

\tikzset{ghost/.style={thick,double,dotted}}

\begin{document}
\thispagestyle{empty}
\vspace*{2em}
\begin{center}
  \textbf{\Large\mathversion{bold}
  BRST quantization of the induced gauge theory  
  \\[5mm] for the Higgs-branch $\CFT_2$}

  \vspace{6em}
  \textrm{\large Bogdan Stefa\'nski, jr.}

  \vspace{4em}
  \begingroup\itshape
  Department of Mathematics,\\
  City St George's, University of London,\\
  Northampton Square, EC1V\,0HB, London, UK
  \par\endgroup

  \vspace{1em}
  \texttt{bogdan.stefanski.1@city.ac.uk}
\end{center}
\vspace{3em}

\begin{abstract}\noindent
We study Witten's induced gauge theory associated with the D1-D5 Higgs-branch $\CFT_2$, in which the non-dynamical vector multiplet becomes reanimated by integrating out the fundamental hypermultiplets. We propose a BRST quantisation of this theory and derive its induced propagators. We show that the ghost sector first contributes at subleading order, that the induced $A\phi^2$ vertex satisfies the Slavnov-Taylor identity and that gauge-invariant correlators are independent of the gauge parameter. Together with our chosen regularization prescription, this gives a perturbative framework for computing observables in the
induced theory.
\end{abstract}

\newpage
\small\tableofcontents
\newpage

\section{Introduction}
\label{sec:intro}

Despite its long history, the $\AdS_3/\CFT_2$ correspondence~\cite{Maldacena:1997re} remains only partially understood away from special points of its moduli space. For the small (4,4) superconformal case considered here it has a 20-dimensional moduli space~\cite{Larsen:1999uk,OhlssonSax:2018hgc,Aharony:2024fid}, sixteen supercharges, and the bulk $\AdS_3\times \Sphere^3\times \Torus^4$ geometries can be supported by Ramond-Ramond (R-R) and Neveu-Schwarz-Neveu-Schwarz (NS-NS) charges
as well as mixed families interpolating between them. The integrable worldsheet string theory is by now well understood~\cite{Babichenko:2009dk,Cagnazzo:2012se}. Symmetry fixes the worldsheet S matrix, including the massless torus
modes~\cite{Borsato:2013qpa,Hoare:2013ida,Hoare:2013lja,Borsato:2014exa,Borsato:2014hja,Lloyd:2014bsa}, and the spectral problem has been formulated
exactly, at least for the RR theory, both via the Quantum Spectral Curve (QSC)~\cite{Cavaglia:2021eqr,Ekhammar:2021pys,Ekhammar:2024kzp} and the Thermodynamic Bethe Ansatz (TBA)~\cite{Frolov:2021bwp}. A TBA has also been proposed for the mixed flux backgrounds~\cite{Frolov:2025tda}. Much less is known about the CFT side. For example, for the R-R background believed to be dual to the origin of the 
Higgs-branch $\CFT_2$ of the D1-D5 system, we lack a weakly-coupled
description for computing observables such as anomalous dimensions of generic
non-BPS operators.~\footnote{The symmetric-product orbifold is
believed to sit at a different point in moduli space.}

The recent QSC solutions~\cite{Ekhammar:2026ykk}
have provided the first results for anomalous dimensions of low-lying operators in the $\algSL(2)$ sector of the R-R theory across the full
range of the integrable coupling $h$. \postarxivchange{At strong coupling, these results agree with the string field theory (SFT) calculation of~\cite{Alday:2026jkn}.\footnote{\postarxivchange{The same work also finds agreement between SFT results for a different family of states and anomalous dimensions extracted from the $\AdS_3$ Virasoro-Shapiro amplitude in a small-curvature expansion~\cite{Chester:2024wnb,Jiang:2025oar}. Together, these comparisons support the mutual consistency of the three approaches.}}} \postarxivchange{At weak coupling, the leading QSC anomalous dimensions} are the same as the energy levels of a nearest-neighbour $s=-\tfrac12$
integrable $\algSL(2)$ spin chain. Any proposed perturbative description of the weakly-coupled
$\CFT_2$ in this neighbourhood
of moduli space should therefore reproduce the 
nearest-neighbour integrable Hamiltonian spectrum in its $\algSL(2)$ sector. 

There is an old proposal for what such a description might be.
Witten~\cite{Witten:1997yu} showed that Higgs-branch states localised near the small-instanton singularity are described by a vector and adjoint hyper
multiplet of the two-dimensional D1-D5 gauge theory, upon dropping the irrelevant super-Yang-Mills kinetic term and integrating out the
fundamental hypermultiplets. In the resulting \textit{induced} gauge
theory the vector multiplet's kinetic and interaction terms are generated entirely by fundamental matter loops, its fields carry geometric rather
than canonical dimensions, and $1/\Nf$ plays the role of a coupling
constant. 

In~\cite{OhlssonSax:2014jtq}, this description was used to compute
the one-loop dilatation operator in the $\algSO(4)$ sector. It was found that the dilatation operator reduced to an integrable nearest-neighbour Hamiltonian of an integrable nearest-neighbour $\algSO(4)$ spin chain, analogous to the $\algSO(6)$ spin-chain in $\mathcal N=4$ super-Yang-Mills~\cite{Minahan:2002ve}. The $\algSO(4)$ calculation can be organised in a 't~Hooft or Veneziano limit, with $\lambda\equiv\Nc/\Nf$ fixed and
small, and with non-planar corrections suppressed by powers of $1/\Nc^2$. The weakly-coupled $\algSL(2)$ dimensions found by the QSC, together with the uniqueness of
the rational nearest-neighbour integrable Hamiltonian that reproduces them, gives us good reason to expect the $\algSL(2)$ spin chain to also follow from the induced theory. 

As a result, a more thorough investigation of the induced theory is timely. In particular, it would be useful to established if the theory, with its non-local effective vertices and induced vector-multiplet kinetic terms can be quantised and gauge-fixed consistently. In this paper, we propose a perturbative BRST quantisation and gauge fixing of the induced theory. Supersymmetry of this theory will be discussed separately in a forthcoming companion paper~\cite{Stefanski:companion}.

This paper is organised as follows. In section~\ref{sec:induced} we review Witten's induced Higgs-branch prescription and state the gauge-fixing and quantisation problem. In Section~\ref{sec:brst-fixing}, using the gauge invariance of determinants, we propose a covariant BRST gauge adapted to the induced kernels of the gauge boson. In section~\ref{sec:brst-rules} we propose a
Double Dimensional Regularisation (DDR) prescription, and use it to compute the quadratic vector-multiplet action, gauge-fixing terms and determine the gauge and ghost propagators. The family of gauges introduced is parametrised by $\xi\in [0,1]$, and for $\xi=1$ the gauge propagator is proportional to $g_{\mu\nu}$. In section~\ref{sec:brst-vertex-identity}, we show that ghost contributions enter at subleading order, we compute the induced $A\phi^2$ vertex and show that it satisfies the Slavnov-Taylor identity at leading order. Section~\ref{sec:evanescent} explains the need for the ordered two-dimensional
limit. We present our conclusions in section~\ref{sec:brst-consistency} More detailed calculations are relegated to the Appendices.

Our results show how the induced theory can be treated with conventional
perturbative methods. Diagrams for gauge-invariant correlators can be
computed and are independent of gauge-fixing parameters. This induced
formulation of the origin of the Higgs-branch $\CFT_2$ is particularly
useful for holography because planar colour contractions can be selected
before performing loop integrals. Just as in $\superN=4$ super Yang--Mills
or in ABJM theory, the perturbation series is arranged in terms of
$\lambda\equiv\Nc/\Nf$, with non-planar diagrams suppressed by inverse
powers of $N_c$.

\section{The induced theory}
\label{sec:induced}

In this section we review Witten's~\cite{Witten:1997yu} induced gauge theory description of the Higgs-branch, and motivate why the induced gauge field kinetic term requires gauge-fixing. This theory was subsequently studied by Aharony and Berkooz~\cite{Aharony:1999dw}, and in the integrable context in~\cite{OhlssonSax:2014jtq}. We use the component Lagrangians and conventions of~\cite{OhlssonSax:2014jtq}, summarised in appendix~\ref{app:microscopic}. The computations in this paper use Euclidean fields. Their continuation from the Lorentzian actions~\cite{OhlssonSax:2014jtq} is also described in appendix~\ref{app:microscopic}.

\subsection{The UV D1-D5 gauge theory and the Higgs branch $\CFT_2$}
We consider $\Nf$ D5-branes stretching along $012345$ and $\Nc$
D1-branes extending along $01$, with the directions $2345$ compactified on a
four-torus~\cite{Maldacena:1997re}. This configuration preserves
$\superN=(4,4)$ supersymmetry in two dimensions. The low-energy modes of the D1-D1 strings give a
 $\grp{U}(\Nc)$ vector multiplet and an adjoint hypermultiplet, the D1-D5 strings give $\Nf$
fundamental hypermultiplets, and the D5-D5 strings decouple in the large volume limit of $\Torus^4$, with $\grp{U}(\Nf)$ a global symmetry. We keep the overall $\grp{U}(1)$ vector multiplet, which couples to the fundamental
hypermultiplets. On the other hand, the trace of the adjoint hypermultiplet is a free
centre-of-mass hypermultiplet and as such decouples.

The action of the two-dimensional multiplets can be found by dimensional reduction from six
dimensions. The vector multiplet contains the gauge field $A_\mu$, four
scalars $\phi^{\alpha\dot\alpha}$ transforming as a vector of the two-dimensional R-symmetry
$\algSO(4)=\algSU(2)_{\sL}\oplus\algSU(2)_{\sR}$ obtained from rotations in $6789$, the fermions $\psi_{\sL}^{\alpha\dot a}$ and
$\psi_{\sR}^{\dot\alpha\dot a}$, and a triplet of auxiliary fields $D^{\dot a\dot b}$. Here $\algSU(2)_\bullet$ is the
six-dimensional R-symmetry, whose doublet carries indices $\dot a$. The fundamental hypermultiplet has complex
scalars $H^{\dot a}$, fermions $\lambda_{\sL}^{\dot\alpha}$ and
$\lambda_{\sR}^{\alpha}$, as well as auxiliary fields $F^{\dot a}$. The adjoint
hypermultiplet consists of scalars $T^{a\dot a}$,  fermions $\chi_{\sL}^{\dot\alpha a}$ and 
$\chi_{\sR}^{\alpha a}$ and the auxiliary $G^{a\dot a}$. The index $a$ is a further
global $\algSU(2)_\circ$ under which only these fields transform.

The ultraviolet (UV) Lagrangian, is fully determined by supersymmetry and takes the form
\begin{equation}\label{eq:uv-total}
\mathcal L_{\rm UV}(\phi,H,T)
=\frac{1}{g_{\rm YM}^2}\,\mathcal L_V(\phi)
+\mathcal L_H(H,\phi)+\mathcal L_T(T,\phi),
\end{equation}
The complete coupling-constant dependence is explicitly given above.
The hypermultiplets couple to the gauge field through representation-appropriate covariant derivatives
with the remaining terms fixed by sypersymmetry. In two dimensions $g_{\rm YM}$ has dimensions of mass and flows in the infrared. Witten argued that in the IR $\mathcal L_V$ should be dropped since it is irrelevant, as can be seen by its $g_{\rm YM}^{-2}$ prefactor, and that the IR Lagrangian is simply
\begin{equation}\label{eq:ir-lag}
\mathcal L_{\rm IR}(\phi,H,T)=\mathcal L_H(H,\phi)+\mathcal L_T(T,\phi)\,.
\end{equation}
This is conformal provided the vector-multiplet fields are assigned \textit{geometric} dimensions
\begin{equation}\label{eq:geom-dims}
\dim(A)=1,\qquad \dim(\phi)=1,\qquad
\dim(\psi)=\tfrac32,\qquad \dim(D)=2\,,
\end{equation}
with the hypermultiplet fields keeping their canonical dimensions. The explicit expressions for $\mathcal L_H(H,\phi)$ and $\mathcal L_T(T,\phi)$ are given in \eqref{eq:uv-fund} and \eqref{eq:uv-adj}. It is easy to check that with these geometric dimensions all terms in $\mathcal L_{\rm IR}(\phi,H,T)$ are classically marginal.

\subsection{The induced action}
Away from its origin, the Higgs branch $\CFT_2$ is a sigma model whose target space is a smooth hyper-K\"ahler manifold. This sigma model can be obtained from~\eqref{eq:ir-lag} by integrating out the (non-dynamical) vector multiplet. At the origin the
instanton size shrinks to zero and the metric is singular. Witten proposed that states localised
there are instead 
described by an \textit{induced} gauge theory obtained from~\eqref{eq:ir-lag} by integrating out the
fundamental hypermultiplets, together with the adjoint hyper-multiplet
\begin{equation}\label{eq:path-int}
\int\!\mathcal D\phi\,\mathcal DT\,\mathcal DH\;
e^{-\int d^2x\,\mathcal L_{\rm IR}}
=\int\!\mathcal D\phi\,\mathcal DT\;
e^{-\int d^2x\,\bigl(\Nf\mathcal L_{\rm eff}(\phi)
+\mathcal L_T(T,\phi)\bigr)} \,.
\end{equation}
The overall factor of $\Nf$ on the right hand side comes from the flavours running in the loop.
Since adjoint and fundamental hypermultiplets do not couple to each other, $\mathcal L_T$ remains the same.
The vector multiplet, on the other hand, acquires new kinetic terms generated entirely by the
fundamental loop, as well as an infinite set of nonlocal interactions. The vector multiplet kinetic term in particular must lead to propagators whose form is fixed by conformal invariance~\eqref{eq:geom-dims}. For example, for scalars
\begin{equation}\label{eq:phi-2pt}
\bigl\langle\phi^{\alpha\dot\alpha}(x)\,\phi^{\beta\dot\beta}(y)\bigr\rangle
=\frac{C\,\epsilon^{\alpha\beta}\epsilon^{\dot\alpha\dot\beta}}{|x-y|^2},
\qquad\text{that is}\qquad
\bigl\langle\phi^{\alpha\dot\alpha}(-p)\,\phi^{\beta\dot\beta}(p)\bigr\rangle
=\frac{C'\,\epsilon^{\alpha\beta}\epsilon^{\dot\alpha\dot\beta}}
{(p^2)^{\frac d2-1}}\,.
\end{equation}
Below, we will provide a dimensional-regularization style prescription for how to compute these induced kernels and interactions, in which we will take the theory to live in dimension $d$, with
\begin{equation}
    d\equiv 2+2\omega\,,\qquad \qquad \mbox{with}\qquad 0<\omega<1\,,
\end{equation}
and continue to $d=2$ at the end.\footnote{The range of $\omega$ is determined by requiring the massless scalar bubble used in defining the induced kernel to be convergent, see~\eqref{eq:brst-BdDhat} for more details.} The normalisations $C$  is not fixed by the fields' geometric dimensions  and will be obtained from the
induced kernel directly in appendix~\ref{app:brst-quadratic}.

Following~\cite{OhlssonSax:2014jtq}, fields which were auxiliary or
non-propagating in the UV acquire nonlocal quadratic terms in the
induced description. For the gauge field, the relevant quantity is
the field-strength two-point function. At quadratic order, the
regulated induced kernel gives this correlator nontrivial momentum
dependence, in contrast to the contact correlator of a two-dimensional
Maxwell field. In the conformal
description, the field strength and the triplet $D^{\dot a\dot b}$
have geometric dimension two and provide the singlet and triplet
used in~\cite{OhlssonSax:2014jtq} to match the field content to the
integrable spin-chain representations constructed on the string
side~\cite{OhlssonSax:2011ms}.

Since $\mathcal L_{\rm IR}$ is quadratic in the fields being integrated
out, \eqref{eq:path-int} is exactly a one-loop determinant and $\mathcal L_{\rm eff}$ interactions
 are one-loop diagrams with fundamental fields running in the loop. For example, the
$A\phi\phi$ vertex is a scalar bubble and a fermion triangle and we can compute both using the explicit vector-hyper interactions of $\mathcal L_H(H,\phi)$ shown in~\eqref{eq:uv-fund}. We construct the
$A\phi\phi$ vertex in section~\ref{sec:brst-induced-Aphiphi}, and appendix~\ref{app:dirac-vertex}.

The perturbative expansion is controlled by two parameters: each effective
vertex comes with a factor of $\Nf$, each vector-multiplet propagator has a factor of $1/\Nf$ and colour loops carry the conventional $\Nc$ factor. As was already discussed in~\cite{OhlssonSax:2014jtq}, in the 't Hooft, or more accurately Veneziano, limit this re-arranges itself into an expansion in
\begin{equation}\label{eq:thooft}
\lambda\equiv\frac{\Nc}{\Nf}\,,
\end{equation}
with non-planar contributions 
suppressed by $1/\Nc^2$.  We can therefore perform
perturbative calculations in the weakly-coupled planar limit, taking 
$\Nc,\Nf\to\infty$ with $\lambda$ fixed and small.

\subsection{The gauge-field kernel}
\label{sec:brst-conv}

As in conventional gauge theories, the induced gauge-field kernel is not invertible and requires gauge-fixing. To see this, we write $A_\mu=A^a_\mu T^a$ with Hermitian generators for
$\mathfrak u(\Nc)$, and 
\begin{equation}\label{eq:brst-conv}
\begin{gathered}
[T^a,T^b]=if^{abc}T^c,\qquad \tr(T^aT^b)=T_F\,\delta^{ab},\\
\nabla_\mu H=(\partial_\mu-iA_\mu)H,\qquad
\nabla_\mu X=\partial_\mu X-i[A_\mu,X]\,.
\end{gathered}
\end{equation}
for fundamental and adjoint fields, respectively. Colour indices run
over all $\Nc^2$ generators, and structure constants vanish when indices include the central $\grp{U}(1)$.

The quadratic in $A$ term in $\Nf\mathcal L_{\rm eff}$ is
the vacuum polarisation of the fundamental loop. Even without evaluating it, we note that current conservation
requires the kernel to be transverse, just as in 
Yang-Mills or Maxwell theory. In other words, the induced kernel has to have the same transverse projector
\begin{equation}
P_{\mu\nu}=g_{\mu\nu}-\frac{p_\mu p_\nu}{p^2}\,,
\end{equation}
as in those familiar theories. This has to be multiplied by a scalar function of momentum that reflects the  geometric scaling of $A_\mu$. 
Since $P_{\mu\nu}$ is dimensionless, the geometric dimension of $A_\mu$ fixes the kernel's remaining momentum dependence to $(p^2)^\omega$ in $d=2+2\omega$. As we show in section~\ref{sec:brst-induced-kernels} and appendix~\ref{app:brst-quadratic}, the explicit loop calculation shows that 
\begin{equation}\label{eq:brst-PiA}
\Gamma^{(2)}_{AA}{}^{ab}_{\mu\nu}(p)
=\delta^{ab}\,K_A(p^2)\Bigl(g_{\mu\nu}-\frac{p_\mu p_\nu}{p^2}\Bigr),
\qquad
K_A(p^2)=C_A\,\Nf\,(p^2)^{\,\omega}\big|_{D\to d}\,,
\end{equation}
where $C_A$ is a momentum-independent coefficient of the induced
kernel. The obstruction to inverting $\Gamma^{(2)}_{AA}$ is therefore the familiar one since $P_{\mu\nu}$ annihilates a longitudinal gauge fluctuation, no matter 
what scalar function multiplies it. As in Yang-Mills or Maxwell theory, gauge-fixing is required to provide the missing longitudinal term needed to invert $\Gamma^{(2)}_{AA}$. The construction we present below will do that in a way that is also adapted to the
induced factor $(p^2)^\omega$.

\section{BRST gauge fixing}
\label{sec:brst-fixing}

In this section we gauge fix the the induced theory and show how this allows us to invert the gauge-field kinetic
operator. Since theory is non-chiral with respect to the
gauge group $\grp{U}(N_c)$, it therefore
has no gauge anomaly. Hence, integrating out the fundamental hypermultiplets preserves the gauge symmetry.\footnote{The R-symmetries $\algSU(2)_{\sL,\sR}$ have 't~Hooft anomalies, but are not gauged.}
As a result, we can apply the BRST construction to the induced theory. As we will see when doing this, it will be convenient to choose the gauge-fixing term to have the same momentum dependence as the induced kinetic term, in order to not introduce any dimensionful gauge-fixing parameters.

We begin with the conventional BRST variation $\delta=[Q,\,\cdot\,\}$. Writing
$c=c^aT^a$ for the ghost, $\bar c$ the antighost and $b$ the Nakanishi--Lautrup field, the transformations are\footnote{In components we have
$\delta A^a_\mu=\partial_\mu c^a+f^{abc}A^b_\mu c^c$ and
$\delta c^a=-\tfrac12f^{abc}c^bc^c$.}
\begin{equation}\label{eq:brst-s}
\delta A_\mu=\nabla_\mu c,\qquad
\delta c=ic^2,\qquad
\delta\bar c=b,\qquad
\delta b=0,\qquad
\delta\phi=i[c,\phi],\qquad
\delta H=icH \,.
\end{equation}
As we review in appendix~\ref{app:brst} (where we also collect our Grassmann and hermiticity conventions), these transformations are nilpotent off shell: $\delta^2=0$.\footnote{The above expressions include the abelian transformations for the overall central $\grp{U}(1)$, so that the construction applies to the full $\grp{U}(\Nc)$ colour algebra.} As a result of gauge invariance of the theory, we have
\begin{equation}
\delta S_{\rm eff}=0\,.
\end{equation}
Hence, for any fermion $\Psi$ of ghost number $-1$ the action
\begin{equation}\label{eq:brst-gauge-fixed-general}
S_\Psi=\Nf S_{\rm eff}+\delta\Psi
\end{equation}
is BRST invariant. As long as the functional measure, and regulators preserve the BRST symmetry, changing $\Psi$ leaves the correlators of BRST-closed
gauge-invariant operators unchanged.\footnote{We work perturbatively around the trivial gauge orbit. Constant gauge transformations need to be divided out separately, as in an ordinary
covariant gauge. We make no claims about the global resolution of the Gribov problem.}

In particular, if $\xi$ is the small parameter of the gauge fermion, then
\begin{equation}
 \partial_\xi S
 =
 \delta\!\left(\partial_\xi\Psi\right)\,,
\end{equation}
and a product $\mathcal O$ of BRST-closed, gauge-invariant operators obeys in the usual way
\begin{equation}
 \partial_\xi\langle\mathcal O\rangle
 =
 -\Bigl\langle
 \mathcal O\,\delta\!\left(\partial_\xi\Psi\right)
 \Bigr\rangle_{\!c}
 =0 \,.
\end{equation}
Physical data such as anomalous
dimensions extracted from such correlators are therefore also
independent of $\xi$. 

As discussed in the previous section, the quadratic kernel for the gauge field is
\begin{equation}
\Gamma^{(2)}_{AA,T}(p)
=C_A\Nf(p^2)^\omega P_{\mu\nu}(p).
\end{equation}
Here $C_A$ is the momentum-independent coefficient introduced in~\eqref{eq:brst-PiA}, and which will be computed in~\eqref{eq:brst-CA-bomega}. We would like to gauge-fix using the standard covariant gauge condition
\begin{equation}
F^a[A]\equiv\partial^\mu A^a_\mu=0\,,
\end{equation}
and so we consider an ansatz for the gauge fermion 
\begin{equation}\label{eq:brst-Psi-general-kernel}
\Psi_{\mathcal K}
=\zeta\Nf\,\tr\!\int d^dx\,
\bar c\,\mathcal K(-\partial^2)
\left(iF[A]+\frac{\xi}{2}b\right)\,,
\end{equation}
where $\mathcal K$ is an unspecified a translationally invariant kernel. Eliminating $b$ would then produce a longitudinal part for the gauge-field kernel
\begin{equation}\label{eq:brst-general-longitudinal}
\Gamma^{(2)}_{AA,L}(p)
=\frac{\zeta\Nf T_F}{\xi}\,
p^2\mathcal K(p^2)L_{\mu\nu}(p)\,,
\end{equation}
where longitudinal projector $L_{\mu\nu}$ is given by
\begin{equation}
L_{\mu\nu}=\frac{p_\mu p_\nu}{p^2}\,.
\end{equation}

It is natural to additionally demand that $\xi$ be dimensionless parameter changing the relative strength of the longitudinal and transverse terms, without introducing another
momentum scale. As a result, we require that their momentum dependence is the same and we fix
the remaining normalisation by requiring equal coefficients at $\xi=1$. This leads to the choice 
\begin{equation}\label{eq:brst-K-selection}
\zeta T_F\,p^2\mathcal K(p^2)
=C_A(p^2)^\omega,
\qquad
\mathcal K(p^2)
=\frac{C_A}{\zeta T_F}(p^2)^{\omega-1}\,.
\end{equation}
Setting $\zeta T_F=C_A$, and denoting this choice of kernel by $K$, we get
\begin{equation}\label{eq:brst-K}
K\equiv(-\partial^2)^{\omega-1}\,,\qquad
\qquad
p^2 K(p^2)=(p^2)^\omega\,,
\end{equation}
and the gauge fermion
\begin{equation}\label{eq:brst-Psi}
\Psi=\zeta\Nf\,\tr\!\int d^dx\,
\bar c\,K\left(i\,\partial\!\cdot\!A+\frac{\xi}{2}\,b\right)\,.
\end{equation}
This choice of kernel follows from the requirements we imposed on top of the BRST symmetry. Other gauge choices are also consistent with the BRST symmetry 
including Landau, axial and dimensionful local covariant
gauges, which we discuss in appendix~\ref{sec:brst-gauge-representatives}. The choice above allows us
to vary $\xi$ without adding a scale and simplifies the comparison of gauge and ghost propagators.

Indeed following the geometric dimensions assignments of the vector multiplet, and our dimensionless gauge fixing parameter, we assign the following dimensions to the ghosts
\begin{equation}\label{eq:brst-dims}
\dim c=0,\qquad\qquad
\dim\bar c=\dim b=2\,.
\end{equation}
Indeed, $\delta A_\mu=\partial_\mu c+\cdots$ gives $\dim c=0$, and since 
$K$ has dimension $d-4$, each term in \eqref{eq:brst-Psi} is marginal. These are the dimensions of the variables in our chosen
representative. We note that the dimensions of 
$(\bar c,b)$ are not invariant and field redefinitions, such as those discussed in appendix~\ref{sec:brst-gauge-representatives} can change this assignment
without changing the BRST cohomology or gauge-invariant correlators.\footnote{The factor of $i$ in \eqref{eq:brst-Psi} is consistent with our hermiticity
conventions. Since $\Psi$ is Grassmann odd,
$(\delta\Psi)^\dagger=-\delta(\Psi^\dagger)$
(appendix~\ref{app:brst}), and a Hermitian gauge-fixed action requires an
anti-Hermitian gauge fermion. With $c,\bar c$ Hermitian and $b$ anti-Hermitian, both terms of \eqref{eq:brst-Psi} are anti-Hermitian. We could equivalently define a Hermitian field $B=ib$ and take the gauge fermion as $\Psi=i\zeta\Nf\,\tr\!\int d^dx\,
\bar cK\left(\partial\!\cdot\!A-\frac{\xi}{2}B\right)$. See, for example~\cite{Becchi:1996an}.}

Applying \eqref{eq:brst-s} with the graded Leibniz rule
\eqref{eq:leibniz} gives
\begin{equation}\label{eq:brst-sPsi}
\delta\Psi=\zeta\Nf\,\tr\!\int d^dx\left[
i\,bK\,\partial\!\cdot\!A
+\frac{\xi}{2}bKb
-i\,\bar cK\,\partial\!\cdot\!\nabla c
\right].
\end{equation}
The equation of motion
$b=-i\,\partial\!\cdot\!A/\xi$ can be used to eliminate $b$ giving
\begin{equation}\label{eq:brst-gf}
S_{\rm gf+gh}
=\zeta\Nf\,\tr\!\int d^dx\left[
\frac{1}{2\xi}
(\partial\!\cdot\!A)K(\partial\!\cdot\!A)
-i\,\bar cK\,\partial\!\cdot\!\nabla c
\right]\,.
\end{equation}

We see explicitly that this is a proper perturbative gauge-fixing since for nonzero Euclidean momentum modes
\begin{equation}
K(p^2)=(p^2)^{\omega-1}>0\,,
\end{equation}
so the first term in \eqref{eq:brst-gf} is positive for
$\xi>0$.\footnote{$K$ can be defined explicitly by expanding fields in Euclidean
Fourier modes and multiplying each mode by
$(p^2)^{\omega-1}$. Since $p^2>0$, this is an ordinary real power. We can then Fourier 
transform back to define a nonlocal operator in position space. For
real-time expressions we analytically continue in the same way for both this factor and the one coming from the induced term.} Additionally, since $K$ is invertible and field-independent
\begin{equation}\label{eq:brst-det-factor}
\ker(KM[A])=\ker M[A],
\qquad\qquad
\det(KM[A])=\det K\,\det M[A]\,.
\end{equation}
As a result, the Fadeev-Popov determinant changes only by a constant compared to its conventional form leaving its perturbative zero-modes and hence gauge-slice unchanged.

Since we set  $\zeta T_F=C_A$, the full quadratic gauge-field kernel is
\begin{equation}\label{eq:brst-good-kernel}
\Gamma^{(2)}_{AA}{}^{ab}_{\mu\nu}(p)
=C_A\Nf\delta^{ab}(p^2)^\omega
\left(P_{\mu\nu}+\frac1\xi L_{\mu\nu}\right)\,,
\end{equation}
which is invertible for $p\ne0$, and at $\xi=1$ it is proportional
to $g_{\mu\nu}$, resembling somewhat the Feynman gauge and validating the choice made in~\cite{OhlssonSax:2014jtq}. The transverse, longitudinal and ghost propagators all come with 
a $1/\Nf$ factor and have the same momentum dependence
$(p^2)^{-\omega}$, as could be expected for a gauge choice that introduced no scale.

We have now constructed a BRST-invariant gauge-fixed action with invertable 
quadratic gauge and ghost kernels.
In the next section we will propose a way to regulate loop integrals in this theory allowing us to explicitly compute the induced quadratic action and propagators  needed for perturbative calculations.

\section{Induced Feynman rules}
\label{sec:brst-rules}

In this section, we explain
how to regulate the loop integrals in this theory, and then collecting the quadratic action and vector-multiplet and ghost propagators, with the 
detailed calculations given in appendix~\ref{app:brst-quadratic}.

\subsection{Double dimensional regularisation}
\label{sec:brst-regulator}
\label{sec:ddr}

A perturbative definition of the induced theory requires a regularisation prescription for its correlation functions which controls both UV divergences and IR singularities
of massless fields in two dimensions. We propose a prescription following the procedure used in appendix~C of~\cite{OhlssonSax:2014jtq} in anomalous-dimension computations. We define the induced propagators in a general dimension $d>2$, and regulate subsequent loop integrals independently in dimension $D$. In this way we are able to remove the
UV regulator \textit{before} taking the two-dimensional limit. We refer to this prescription as double dimensional regularisation (DDR), since it involves $d$ and $D$.

The distinction between the two dimensions is physically important. $d$ determines the conformal weights of the continued theory, equivalently the powers of momentum in the induced propagators. These powers stay
fixed as we vary the loop integration dimension $D$.  We write
\begin{equation}\label{eq:brst-ddr-parameters}
 d=2+2\omega,
 \qquad
 D=d-\varepsilon=2+2\omega-\varepsilon,
 \qquad
 0<\varepsilon<2\omega,
\end{equation}
with $\omega>0$ in a neighbourhood of the physical limit, so that $2<D<d$.\postarxivchange{\footnote{\postarxivchange{More generally, loop integrals involving the adjoint hypermultiplet may first be evaluated in a smaller convergence strip, $2<D<D_*<d$, and continued analytically in $D$ towards $D=d$ before taking the ordered limits below. This refinement is not needed for the explicit calculations considered here.}}} Loop loop momenta are then integrated with measure
\begin{equation}\label{eq:brst-ddr-measure}
 \mu^{d-D}\int\!\frac{d^Dq}{(2\pi)^D}
 =\mu^{\varepsilon}\int\!\frac{d^Dq}{(2\pi)^D} \,,
\end{equation}
and the regulated action may be written as $S_{d,D}=\mu^{D-d}\int d^Dx\,\mathcal L_d$.  The factor of $\mu$ ensures the $d$-dimensional engineering dimensions work out and supplies the renormalisation scale when a pole at $D=d$ is subtracted. Having performed the computation for generic $d$ and $D$, We first remove the UV regulator keeping  $d>2$ fixed. Only then do we take the $d\rightarrow$ limit at the end. In other words, after combining diagrams that contribute to a given  amplitude and subtracting the UV poles at
we take
\begin{equation}\label{eq:brst-ddr-limits}
 \qquad
 \varepsilon\longrightarrow0^+\quad\text{at fixed }\omega>0,
 \qquad\text{followed by}\qquad
 \omega\longrightarrow0^+ \,.
 \qquad
\end{equation}
For example, the residues of subtracted UV poles are retained when they determine anomalous dimensions, and gauge-related diagrams are assembled together before the regulator is removed.

The reason for the DDR procedure can be seen by considering the following integral
\begin{equation}\label{eq:double-dim-example}
 J(D,d;p)
 =\mu^{d-D}\int\!\frac{d^Dq}{(2\pi)^D}\,
 \frac{1}{(q^2)^\omega(q+p)^2}\,,
 \qquad p^2\ne0 \,.
\end{equation}
At large $q$, the radial integrand goes like
$q^{-1-\varepsilon}=q^{D-d-1}$, so UV convergence requires
$D<d$. Near $q=-p$, we instead have a scaling like $q^{2\omega-\varepsilon-1}=q^{D-3}$, so for IR convergence we need
$D>2$. Finally at $q=0$, the radial behaviour is
$q^{1-\varepsilon}$. This is integrable in the neighbourhood of
$\omega=0^+$, which is consistent with the DDR limit, since
\(0<\varepsilon<2\omega<2\).

The order of limits matters because of the soft scalar endpoint
enhancements encountered below. Near a soft massless
fundamental scalar, the radial integral has a pole. Taking the first limit $D\to d$, this becomes
\begin{equation}\label{eq:brst-ddr-soft}
\int_0 dq\;q^{\,2\omega-\varepsilon-1}
=\frac{1}{2\omega-\varepsilon}
 \xrightarrow{\ \varepsilon\to0^+\ }
\frac{1}{2\omega}\,.
\end{equation}
Before the limit, the denominator is $2\omega-\varepsilon$,  afterwards the soft region gives an $\omega^{-1}$ factor, which then can
compensate a zero of an induced propagator in the $d\rightarrow 2$ limit.

We apply DDR to diagrams computed in the induced action, including its BRST-exact gauge-fixing sector. The BRST transformations stay the same, and
we can keep the $d$-dependent form factors while using the same
dimensional prescription for tensor contractions and loop
integrations. As we will show below, once gauge-related diagrams combine before the ordered limits are taken and UV poles are subtracted, the assembled amplitudes satisfy 
Slavnov-Taylor identities as required.

Finally, for clarity, we will distinguish the loop integrations in the induced theory from the fundamental
loops that define the induced action itself. In the latter case we will denote the loop integral dimension as $\widehat D$, and taking $\widehat D\to d$ before forming diagrams from the resulting induced
kernels and vertices. The full construction can be thought of as
\begin{equation}\label{eq:brst-ddr-hierarchy}
 \Gamma_{\rm fund}^{(d,\widehat D)}
 \xrightarrow{\ \widehat D\to d\ }
 S_{\rm ind}^{(d)}
 \xrightarrow{\ \text{loops in }D<d\ }
 \Gamma_{d,D} \,.
\end{equation}

\subsection{The quadratic action}
\label{sec:brst-induced-kernels}

The induced quadratic vector-multiplet come from a single loop of fundamental fields, and diagramatically are
\begin{equation}\label{eq:inducedblobs}
\begin{aligned}
  \begin{tikzpicture}[baseline=-0.5ex]
    \draw [gluon] (-0.6cm,0) -- (0.6cm,0);
    \fill [] (0,0) circle [radius=0.125cm];
  \end{tikzpicture}
  &=
  \begin{tikzpicture}[baseline=-0.5ex]
    \coordinate (v1) at (-0.35cm,0); \coordinate (v2) at (0.35cm,0);
    \draw [gluon] (-0.85cm,0) -- (v1);
    \draw [gluon] (v2) -- (0.85cm,0);
    \draw [hyp scalar,out=60,in=120] (v1) to (v2);
    \draw [hyp scalar,out=300,in=240] (v1) to (v2);
    \fill [] (v1) circle [radius=0.0625cm];
    \fill [] (v2) circle [radius=0.0625cm];
  \end{tikzpicture}
  +
  \begin{tikzpicture}[baseline=-0.5ex]
    \coordinate (v1) at (-0.35cm,0); \coordinate (v2) at (0.35cm,0);
    \draw [gluon] (-0.85cm,0) -- (v1);
    \draw [gluon] (v2) -- (0.85cm,0);
    \draw [hyp fermion,out=60,in=120] (v1) to (v2);
    \draw [hyp fermion,out=300,in=240] (v1) to (v2);
    \fill [] (v1) circle [radius=0.0625cm];
    \fill [] (v2) circle [radius=0.0625cm];
  \end{tikzpicture}
  +
  \begin{tikzpicture}[baseline=-0.5ex]
    \draw [gluon] (-0.6cm,0) -- (0.6cm,0);
    \draw [hyp scalar] (0,0.3cm) circle [radius=0.3cm];
    \fill [] (0,0) circle [radius=0.0625cm];
  \end{tikzpicture}
  \\[4pt]
  \begin{tikzpicture}[baseline=-0.5ex]
    \draw [scalar] (-0.6cm,0) -- (0.6cm,0);
    \fill [] (0,0) circle [radius=0.125cm];
  \end{tikzpicture}
  &=
  \begin{tikzpicture}[baseline=-0.5ex]
    \coordinate (v1) at (-0.35cm,0); \coordinate (v2) at (0.35cm,0);
    \draw [scalar] (-0.85cm,0) -- (v1);
    \draw [scalar] (v2) -- (0.85cm,0);
    \draw [hyp fermion,out=60,in=120] (v1) to (v2);
    \draw [hyp fermion,out=300,in=240] (v1) to (v2);
    \fill [] (v1) circle [radius=0.0625cm];
    \fill [] (v2) circle [radius=0.0625cm];
    \node at (0,0.62cm) {\scriptsize $\lambda_{L}$};
    \node at (0,-0.62cm) {\scriptsize $\lambda_{R}$};
  \end{tikzpicture}
  +
  \begin{tikzpicture}[baseline=-0.5ex]
    \draw [scalar] (-0.6cm,0) -- (0.6cm,0);
    \draw [hyp scalar] (0,0.3cm) circle [radius=0.3cm];
    \fill [] (0,0) circle [radius=0.0625cm];
  \end{tikzpicture}
  \\[4pt]
  \begin{tikzpicture}[baseline=-0.5ex]
    \draw [fermion] (-0.6cm,0) -- (0.6cm,0);
    \fill [] (0,0) circle [radius=0.125cm];
  \end{tikzpicture}
  &=
  \begin{tikzpicture}[baseline=-0.5ex]
    \coordinate (v1) at (-0.35cm,0); \coordinate (v2) at (0.35cm,0);
    \draw [fermion] (-0.85cm,0) -- (v1);
    \draw [fermion] (v2) -- (0.85cm,0);
    \draw [hyp scalar,out=60,in=120] (v1) to (v2);
    \draw [hyp fermion,out=300,in=240] (v1) to (v2);
    \fill [] (v1) circle [radius=0.0625cm];
    \fill [] (v2) circle [radius=0.0625cm];
    \node at (0,0.62cm) {\scriptsize $H$};
    \node at (0,-0.62cm) {\scriptsize $\lambda_{L}$};
  \end{tikzpicture}
  \\[4pt]
  \begin{tikzpicture}[baseline=-0.5ex]
    \draw [aux] (-0.6cm,0) -- (0.6cm,0);
    \fill [] (0,0) circle [radius=0.125cm];
  \end{tikzpicture}
  &=
  \begin{tikzpicture}[baseline=-0.5ex]
    \coordinate (v1) at (-0.35cm,0); \coordinate (v2) at (0.35cm,0);
    \draw [aux] (-0.85cm,0) -- (v1);
    \draw [aux] (v2) -- (0.85cm,0);
    \draw [hyp scalar,out=60,in=120] (v1) to (v2);
    \draw [hyp scalar,out=300,in=240] (v1) to (v2);
    \fill [] (v1) circle [radius=0.0625cm];
    \fill [] (v2) circle [radius=0.0625cm];
  \end{tikzpicture}
\end{aligned}
\end{equation}
The gauge field sum includes an $H$ bubble, two chiral-fermion
bubbles as well as the $H^\dagger A^2 H$ tadpole. The $\phi$ kernel has a mixed chirality fermion loop and the
$H^\dagger\phi^2 H$ tadpole. The $\psi_{\sR}$ kernel is drawn above and the $\psi_{\sL}$ kernel instead involves a $H$-$\lambda_{\sR}$ bubble. Finally, the $D$ kernel is an $H$ bubble. The fundamental propagators and vertices
are given in \eqref{eq:brst-fund-propagator-residues} and \eqref{eq:brst-H-gauge-rules}.

The detailed evaluation of these diagrams using the DDR prescription is given in Appendix~\ref{app:component-kernels} and leads to the following quadratic action.\footnote{\label{fn:brst-scalar-metric}
Our scalar convention is $\phi_{\alpha\dot\alpha}\phi^{\alpha\dot\alpha}
=2\phi_I\phi_I$, with $I=1,\ldots,4$. Thus
$\tfrac12\phi_I K_\phi\phi_I
=\tfrac14\phi_{\alpha\dot\alpha}K_\phi\phi^{\alpha\dot\alpha}$:
the real-component covariance has coefficient $K_\phi^{-1}$ and
the bispinor covariance has coefficient $2K_\phi^{-1}$.}
\begin{equation}\label{eq:brst-vector-quadratic-functional}
\begin{aligned}
 \Gamma_{\rm ind}^{(2)}
 =\int\!\frac{d^dp}{(2\pi)^d}\Bigl[
 &\tfrac12A^a_\mu(-p)K_A^{ab,\mu\nu}(p)A^b_\nu(p)
 +\tfrac14\phi^a_{\alpha\dot\alpha}(-p)
 K^{ab;\alpha\dot\alpha}_{\phi\;\beta\dot\beta}(p)
 \phi^{b\beta\dot\beta}(p)
\\
 &+\bar\psi^a_{\sR\,\dot\alpha\dot a}(-p)
 K_{\psi_{\sR}}^{ab}(p)\psi_{\sR}^{b\dot\alpha\dot a}(p)
 +\bar\psi^a_{\sL\,\alpha\dot a}(-p)
 K_{\psi_{\sL}}^{ab}(p)\psi_{\sL}^{b\alpha\dot a}(p)
\\
 &+\frac12 D^{ar}(-p)K_D^{ab;rs}(p)D^{bs}(p)\Bigr]\,.
\end{aligned}
\end{equation}
\noindent The bosonic kernels are\nopagebreak
\begin{equation}\label{eq:kernelsbos}
\begin{aligned}
 K_A^{ab,\mu\nu}(p)
 &=2\Nf T_F\delta^{ab}
 (p^2\delta^{\mu\nu}-p^\mu p^\nu)B_{d,\widehat D}(p),
\\
 K^{ab;\alpha\dot\alpha}_{\phi\;\beta\dot\beta}(p)
 &=2\Nf T_F\delta^{ab}\delta^\alpha{}_\beta\delta^{\dot\alpha}{}_{\dot\beta}
 p^2B_{d,\widehat D}(p),
\\
 K_D^{ab;rs}(p)
 &=2\Nf T_F\delta^{ab}\delta^{rs}
 B_{d,\widehat D}(p),
\end{aligned}
\end{equation}
and  fermionic kernels\footnote{The kernel on independent components is $2K_\psi$, as explained in
appendix~\ref{app:fermion-normalization}.}
\begin{equation}\label{eq:kernelsferm}
 K_{\psi_{\sR}}^{ab}(p)
 =2i\Nf T_F\delta^{ab}p_-B_{d,\widehat D}(p)\,,
 \qquad\qquad
 K_{\psi_{\sL}}^{ab}(p)
 =2i\Nf T_F\delta^{ab}p_+B_{d,\widehat D}(p)\,.
\end{equation}
Above, $B_{d,\widehat D}$ is the DDR expression for the fundamental bubble defined  in
\eqref{eq:brst-BdDhat}.

The common form factor in the quadratic kernels makes it useful to
collect them into a single functional. Restricting the external fields
and derivatives to the physical two-dimensional spacetime, we retain
the dependence on $d$ and $\widehat D$ in the even scalar form factor
\begin{equation}\label{eq:brst-common-form-factor}
 \mathcal B(p^2)=2\Nf B_{d,\widehat D}(p),
 \qquad \mathcal B\equiv\mathcal B(-\partial^2).
\end{equation}
No limit of the bubble integral is needed for this rewriting.
With colour indices suppressed, the Euclidean quadratic
action~\eqref{eq:brst-vector-quadratic-functional} becomes\footnote{Actions with fractional Laplacians~\cite{Kwasnicki:20157356}
 appear in many settings, such as~\cite{Heydeman:2020ijz}
 and references therein.}
\begin{equation}\label{eq:brst-quadratic-action-2d}
\begin{aligned}
 \Gamma_{\rm ind}^{(2)}=T_F\int d^2x\,\Bigl[&
 \tfrac12 F_{12}\mathcal B F_{12}
 +\tfrac14\partial_\mu\phi_{\alpha\dot\alpha}
 \mathcal B\partial_\mu\phi^{\alpha\dot\alpha}
 +\tfrac12D^r\mathcal BD^r\\
 &+\bar\psi_{\sL\,\alpha\dot a}
\mathcal B\partial_+\psi_{\sL}^{\alpha\dot a}
 +\bar\psi_{\sR\,\dot\alpha\dot a}
  \mathcal B\partial_-\psi_{\sR}^{\dot\alpha\dot a}\Bigr]\,.
\end{aligned}
\end{equation}
Here $F_{12}=\partial_1A_2-\partial_2A_1$ and
$\partial_\pm=\partial_1\pm i\partial_2$.\footnote{Here $F_{12}$ is the linearised non-Abelian field
strength. The commutator term $-i[A_1,A_2]$ and the gauge-connection
terms in covariant derivatives first enter the action at cubic order.
The simple form of~\eqref{eq:brst-quadratic-action-2d} suggests a natural
way to organise the higher interactions: restore the full field
strength and replace all derivatives by covariant derivatives,
including $\mathcal B(-\partial^2)\to\mathcal B(-\nabla^2)$ acting on
adjoint fields. This gives a gauge-invariant completion, but it is not
unique: different operator orderings and additional gauge-invariant
terms can agree at quadratic order; see~\cite{Ilderton:2008gauge} for a
discussion of this freedom in nonlocal actions. The longitudinal
bubble contribution to the $A\phi^2$ vertex
in~\eqref{eq:brst-Aphiphi-d} is fixed by the Ward identity and is
reproduced by this completion. The completion also generates transverse
bubble terms, so the required transverse correction is the triangle
contribution calculated in Appendix~\ref{app:dirac-vertex} minus those
transverse bubble terms. More generally, the induced action can be
organised as this natural covariant completion plus further
gauge-invariant interactions determined by the fundamental
determinant. I am grateful to Chris Hull for discussions about the form of the kinetic action.}
The fermion terms contain the full component sums. The fields $D^r$,
with $r=1,2,3$, are the three components of the original auxiliary
$\algSU(2)_\bullet$ triplet,
$D^{\dot a\dot b}=D^r(\sigma^r)^{\dot a\dot b}$, in the
Pauli-matrix normalization $\tr(\sigma^r\sigma^s)=2\delta^{rs}$.

Next, we take the defining limit $\widehat D\to d=2+2\omega$. The bubble then becomes
\begin{equation}\label{eq:brst-bomega}
 B_{d,d}(p)=b_\omega(p^2)^{\omega-1},
 \qquad\qquad
 b_\omega=(4\pi)^{-1-\omega}
 \frac{\Gamma(1-\omega)\Gamma(\omega)^2}{\Gamma(2\omega)} \,.
\end{equation}
The small-$\omega$ behaviour is
\begin{equation}\label{eq:brst-bomega-limit}
 b_\omega
 =\frac{1}{2\pi\omega}
 +\frac{\gamma_{\rm E}-\log(4\pi)}{2\pi}
 +\Ord(\omega),
 \qquad\qquad
 \lim_{\omega\to0^+}\omega b_\omega=\frac{1}{2\pi}\,.
\end{equation}
Finally, inverting the quadratic form gives the
following ungauge-fixed propagators:
\begin{equation}\label{eq:brst-vector-propagators}
\begin{aligned}
 \langle A^a_\mu(-p)A^b_\nu(p)\rangle_{\rm T}
 &=\frac{\delta^{ab}}{2\Nf T_Fb_\omega}
 (p^2)^{-\omega}
\left(g_{\mu\nu}-\frac{p_\mu p_\nu}{p^2}\right)\,,
 \\
 \langle\phi^{aI}(-p)\phi^{bJ}(p)\rangle
 &=\frac{\delta^{ab}\delta^{IJ}}{2\Nf T_Fb_\omega}
 (p^2)^{-\omega}\,,
\\
 \langle\psi^a_{\sR}(p)\bar\psi^b_{\sR}(-p)\rangle
 &=-\frac{i\delta^{ab}}{4\Nf T_Fb_\omega}
 \frac{(p^2)^{1-\omega}}{p_-}
 =-\frac{i\delta^{ab}}{4\Nf T_Fb_\omega}
 p_+(p^2)^{-\omega}\,,
\\
 \langle\psi^a_{\sL}(p)\bar\psi^b_{\sL}(-p)\rangle
 &=-\frac{i\delta^{ab}}{4\Nf T_Fb_\omega}
 \frac{(p^2)^{1-\omega}}{p_+}
 =-\frac{i\delta^{ab}}{4\Nf T_Fb_\omega}
 p_-(p^2)^{-\omega}\,,
\\
 \langle D^{ar}(-p)D^{bs}(p)\rangle
 &=\frac{\delta^{ab}\delta^{rs}}{2\Nf T_Fb_\omega}
 (p^2)^{1-\omega}.
\end{aligned}
\end{equation}
The $A$ kernel is transverse and in the next sub-section we complete it via gauge-fixing.
We can now read-off the normalization of the two-point functions, which in the notation of \eqref{eq:brst-PiA} gives
\begin{equation}\label{eq:brst-CA-bomega}
 C_A=2T_Fb_\omega \,.
\end{equation}

\subsection{Gauge and ghost propagators}

The gauge-finxing procedure in section~\ref{sec:brst-fixing} uses the kernel $K=(-\partial^2)^{\omega-1}$. Matching the gauge fermion's longitudinal kernel to the transverse propagator~\eqref{eq:kernelsbos} fixes
\begin{equation}\label{eq:brst-zeta-bomega}
 \zeta=2b_\omega,
 \qquad
 \zeta T_F=C_A=2T_Fb_\omega \,.
\end{equation}
As a result the complete two-point function can be written as
\begin{equation}\label{eq:brst-kernel}
\Gamma^{(2)}_{AA}{}^{ab}_{\mu\nu}(p)
=2\Nf T_Fb_\omega\delta^{ab}(p^2)^\omega
\left(P_{\mu\nu}+\frac1\xi L_{\mu\nu}\right).
\end{equation}
Since \(P\) and \(L\) are orthogonal projectors, we can invert trivially to get
\begin{equation}\label{eq:brst-AA}
\langle A^a_\mu(-p)A^b_\nu(p)\rangle
=\frac{\delta^{ab}}{2\Nf T_Fb_\omega}(p^2)^{-\omega}
\left(P_{\mu\nu}+\xi L_{\mu\nu}\right) \,.
\end{equation}
For $\xi=1$ this is proportional to $\delta_{\mu\nu}$ and produces a Feynman-gaugelike propagator used in~\cite{OhlssonSax:2014jtq}.

We similarly find the ghost propagator 
\begin{equation}\label{eq:brst-ghprop}
 \langle c^a(p)\bar c^b(-p)\rangle
 =-\frac{i\delta^{ab}}{2\Nf T_Fb_\omega}(p^2)^{-\omega},
\end{equation}
and the ghost-gauge interaction is
\begin{equation}\label{eq:brst-ghint}
 S_{\bar cAc}
 =-2i\Nf T_Fb_\omega f^{abc}
 \int d^dx\,
 \bar c^{\,a}K\,\partial^\mu(A_\mu^b c^c)\,.
\end{equation}
The corresponding ghost-gluon vertex is
\begin{equation}\label{eq:brst-ghvertex}
 V_\mu^{abc}(\bar p)
 =2\Nf T_Fb_\omega f^{abc}
 (\bar p^2)^{\omega-1}\bar p_\mu,
\end{equation}
where \(\bar p\) is the anti-ghost momentum. 

In appendix~\ref{sec:brst-gauge-representatives} we discuss other ghost variables which lead to a local Faddeev-Popov operator, and Landau and local-covariant gauges. The advantage of the procedure outlined in the body of the paper is that the transverse, longitudinal and ghost kernels have the same weights and $\xi$ is dimensionless. This allows us to treat all three propagators in the same way when counting powers of the expansion parameters.

\section{Ghosts and the induced $A\phi^2$ vertex}
\label{sec:brst-vertex-identity}

In this section we show that BRST ghost contribute at sub-leading order in the induced theory, thus leaving the kernels computed in the previous section unaffected. We then determine
of the leading order Slavnov-Taylor identity and show that it is satisfied in the case of the induced $A\phi\phi$ vertex and the $\phi$ two-point function.

\subsection{Ghost contributions}
\label{sec:brst-counting}

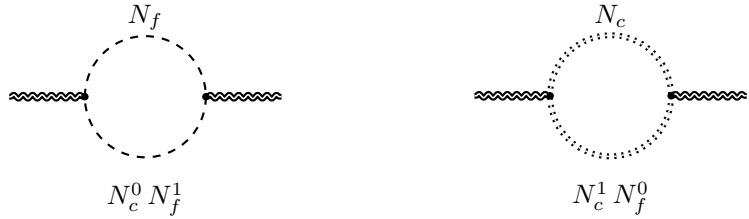
\begin{figure}[htb]
\centering
\begin{tikzpicture}[baseline=(current bounding box.center)]
  \coordinate (l)  at (-1.8,0);
  \coordinate (v1) at (-0.8,0);
  \coordinate (v2) at ( 0.8,0);
  \coordinate (r)  at ( 1.8,0);
  \draw [gluon] (l) -- (v1);
  \draw [gluon] (v2) -- (r);
  \draw [hyp fermion] (v1) arc (180:0:0.8);
  \draw [hyp fermion] (v2) arc (0:-180:0.8);
  \fill (v1) circle (1.4pt);
  \fill (v2) circle (1.4pt);
  \node at (0,1.05) {\footnotesize $\Nf$};
  \node at (0,-1.35) {\footnotesize $\Nc^{0}\,\Nf^{1}$};
\end{tikzpicture}
\hspace{2.2cm}
\begin{tikzpicture}[baseline=(current bounding box.center)]
  \coordinate (l)  at (-1.8,0);
  \coordinate (v1) at (-0.8,0);
  \coordinate (v2) at ( 0.8,0);
  \coordinate (r)  at ( 1.8,0);
  \draw [gluon] (l) -- (v1);
  \draw [gluon] (v2) -- (r);
  \draw [ghost] (v1) arc (180:0:0.8);
  \draw [ghost] (v2) arc (0:-180:0.8);
  \fill (v1) circle (1.4pt);
  \fill (v2) circle (1.4pt);
  \node at (0,1.05) {\footnotesize $\Nc$};
  \node at (0,-1.35) {\footnotesize $\Nc^{1}\,\Nf^{0}$};
\end{tikzpicture}
\caption{A fundamental-matter loop and a ghost loop contributing to the
non-abelian gauge two-point function. The single lines on the left denote
fundamental fields. Their free flavour sum makes this diagram of order
$\Nf$. The double lines on the right denote adjoint ghosts, whose closed
colour loop makes the diagram of order $\Nc$.
The ratio of the two is $\Nc/\Nf=\lambda$.}
\label{fig:ghost-vs-matter}
\end{figure}

Comparing a fundamental-matter loop to a ghost loop is straightforward.\footnote{The overall $\mathfrak u(1)$ ghosts are free in our chosen gauge.} The gauge field and $\phi$ kinetic terms come from
matter loops, giving an overall factor of $\Nf$. Ghosts come from gauge-fixing, and their kinetic term normalisation $\zeta\Nf T_F$ also multiplies their interaction vertex. As a result, it cancels around any closed ghost loop. We are left with a colour trace compared to the fundamental matter loop. For two such legs, the ghosts therefore contribute
$f^{acd}f^{bcd}=2T_F\Nc\delta^{ab}$ for external traceless indices $a,b$. on the other hand the fundamental matter loop gives $\Nf T_F\delta^{ab}$.  Thus
\begin{equation}\label{eq:brst-lemma}
\frac{\text{closed ghost loop with $V$ gluon legs}}
{\text{fundamental-matter loop with $V$ gluon legs}}
\ \sim\ {\frac{2T_F\Nc}{\Nf T_F}}\ \propto\ \lambda \,.
\end{equation}
As a result,  ghost loops are $\lambda$-suppressed compared to induced vertices, as is the case in the gluon self-energy or
the induced $A^3$ vertex. 
An induced-gluon exchange between a ghost or scalar line
starts at relative $\Ord(\lambda)$, because it contains an induced
propagator $\Nf^{-1}$ and an adjoint colour contraction $\Nc$. As we saw above, a ghost loop inside
these kernels would therefore first enters at relative
$\Ord(\lambda^2)$. Similarly,  closed gauge field loops have the same suppression as ghost loops, since both are adjoint loops. Figure~\ref{fig:ghost-vs-matter} summarises the above argument diagramatically. 
For diagrams with internal gauge lines, this argument uses
our chosen normalisation of the gauge-fixing term. 
Matching the gauge-fixing term with the transverse kernel ensures the full gauge propagator scales as $1/\Nf$.

\subsection{The induced $A\phi^2$ vertex and its Slavnov-Taylor identity}
\label{sec:brst-st}
\label{sec:brst-induced-Aphiphi}

The induced action $\Gamma_{\rm ind}=\Nf S_{\rm eff}$ is gauge invariant
\begin{equation}
 e^{-\Gamma_{\rm ind}[A^U,\phi^U]}
 =\int\!\mathcal DH^U\,
 e^{-S_H[H^U;A^U,\phi^U]}\\
 =\int\!\mathcal DH\,
 e^{-S_H[H;A,\phi]}
 =e^{-\Gamma_{\rm ind}[A,\phi]}\,,
\end{equation}
with $U(x)$ a gauge transformation and $S_H=\int\mathcal L_H$.\footnote{This
follows from the gauge invariance of the action, DDR prescription and an anomaly-free measure.} As a result, the condition $\delta\Gamma_{\rm ind}=0$. The induced action's $\phi^2$ term has a gauge variation proportional to $\alpha\phi^2$, where $\alpha$ is the gauge-transformation parameter. Varying $A$ in the induced action's cubic term $A\phi^2$ also leads to a term proportional to $\alpha\phi^2$
Since the induced action is gauge invariant, these
contributions must cancel. Working to leading order in $lambda$, as in the previous section, both contributions come from the one-loop diagrams of the integrated-out fundamental hypermultiplet. The common real scalar component index $I$ is suppressed in the
following expressions. Explicitly,
\begin{equation}
\begin{aligned}
 \Gamma_2&=\frac12\int \frac{d^dk}{(2\pi)^d}
 \phi^r(-k)K_\phi(k^2;d)\phi^r(k)\,,\\
 \Gamma_3&=\frac12\int \frac{d^dp}{(2\pi)^d}\int \frac{d^dq}{(2\pi)^d}
 A^a_\mu(p)\phi^b(q)\Gamma^{a\mu}_{bc}(q,p;d)\phi^c(-b)\,,
\end{aligned}
\end{equation}
where $b=q+p$, $K_\phi$ is given in~\eqref{eq:kernelsbos}, and $\Gamma^{a\mu}_{bc}$ is a third-order derivative of the action with its momentum delta-function removed.
Using the scalar metric in footnote~\ref{fn:brst-scalar-metric}, the same cubic
action in bispinor notation is
\begin{equation}\label{eq:brst-Aphiphi-scalar-metric}
 \Gamma_3=\frac14\int_{p,q}
 A^a_\mu(p)\phi^b_{\alpha\dot\alpha}(q)
 \Gamma^{a\mu}_{bc}(q,p;d)\phi^{c\alpha\dot\alpha}(-b).
\end{equation}
Taking $p$ and $q$ as incoming and $b$ as outgoing, and Fourier transforming to momentum space, the gauge variations of $\Gamma_2$ and $\Gamma_3$ contributing to $\alpha\phi^2$ are
\begin{equation}
\begin{aligned}
 \delta\Gamma_2&=\frac12\int_{p,q}
 \alpha^a(p)\phi^b(q)\phi^c(-b)\,
 f^{abc}\left[K_\phi(q^2;d)-K_\phi(b^2;d)\right]\,,\\
 \left.\delta\Gamma_3\right|_{A=0}
 &=\frac12\int_{p,q}
 \alpha^a(p)\phi^b(q)\phi^c(-b)\,
 ip_\mu\Gamma^{a\mu}_{bc}(q,p;d)\,.
\end{aligned}
\end{equation}
Gauge invariance requires their sum to vanish, giving the leading-order Slavnov-Taylor identity
\begin{equation}\label{eq:brst-leading-ST}
 i p_\mu\Gamma^{a\mu}_{bc}(q,p;d)
 =f^{abc}\left[K_\phi(b^2;d)-K_\phi(q^2;d)\right] \,.
\end{equation}
As we discussed in section~\ref{sec:brst-counting}, ghost dressing and scattering-kernel corrections will contribute at subleading orders. In Appendix~\ref{app:dirac-vertex}, we computed $\Gamma^{a\mu}_{bc}$ to leading order and found
\begin{equation}\label{eq:brst-Aphiphi-d}
\begin{aligned}
 \Gamma^{a\mu}_{bc}(q,p;d)= -2i\Nf T_Ff^{abc}\Bigl\{&
 \frac{p^\mu}{p^2}b_\omega
 \left[(b^2)^\omega-(q^2)^\omega\right]
 \\ &
 +8\pi(d-2)C_{d+2}(p^2,q^2,b^2)
 \left[q^\mu-\frac{p\cdot q}{p^2}p^\mu\right]\Bigr\}\,,
\end{aligned}
\end{equation}
where $C_{d+2}$ is a symmetric simplex integral over Feynman parameters given in~\eqref{eq:dirac-triangle}. Above, external momenta and gauge-field index are taken to lie in the physical plane. To verify that the Slavnov-Taylor identity is satisfied, we contract $\Gamma^{a\mu}_{bc}$ with $ip_\mu$, which removes the transverse term on the second line of~\eqref{eq:brst-Aphiphi-d}. The terms on the right hand side of the first line gives precisely the difference of scalar kernels in~\eqref{eq:brst-leading-ST}. We note that in the $d\rightarrow 2$ limit 
\begin{equation}
 b_\omega\left[(b^2)^\omega-(q^2)^\omega\right]
 \xrightarrow{\ \omega\to0\ }
 \frac{1}{2\pi}\log\frac{b^2}{q^2}\,,
\end{equation}
and $C_{d+2}\to C_4<\infty$ for generic momenta. In this limit the transverse term decouples from the vertex completely leaving the finite expression
\begin{equation}\label{eq:brst-Aphiphi-2d}
 \Gamma^{a\mu}_{bc}(q,p)\big|_{d=2}
 =-\frac{i\Nf T_Ff^{abc}}{\pi}\,
 \frac{p^\mu}{p^2}\log\frac{b^2}{q^2} \,.
\end{equation}
This form of the vertex can substantially simplify further diagrammatic calculations, as long as the omitted evanescent~\cite{Buras:1989xd,Dugan:1990df} terms vanish after any remaining loop integrals. As we discuss below,  it is possible for an integration pole to compensate for this zero and lead to a finite expression, in which case the complete but more complicated vertex is needed.

\section{The two-dimensional limit and evanescence}
\label{sec:evanescent}

The induced propagators~\eqref{eq:brst-vector-propagators} vanish in the $d\rightarrow 2$ limit, because of the $b_\omega$ pole~\eqref{eq:brst-bomega-limit}, since\footnote{In comparison, the Schwinger model retains its kinetic $F^2$ term. The vacuum polarisation, analogous to the integrating out of the fundamental hypers here, modifies the inverse propagator generating a photon mass. Here the \textit{entire} transverse kinetic operator is obtained from the one loop diagram.}
\begin{equation}\label{eq:brst-evanescent-propagator}
 \frac{1}{2\Nf T_Fb_\omega}(p^2)^{-\omega}
 =\frac{\pi\omega}{\Nf T_F}
 \left[1+\Ord(\omega)\right]\,.
\end{equation}
As a result, the internal induced lines of any graph come with an explicit factor of $\omega^{I_{\rm ind}}$, where $I_{\rm ind}$ counts those lines. As we saw above, the $A\phi^2$ vertex is finite in the $d\rightarrow 2$ limit because the pole canceled, leaving only $\ln(b^2/q^2)$, with its transverse part vanishing. This vertex therefore does not give rise to inverse powers of $\omega$ to
compensate for the propagator zeros.

More generally, whether a diagram with internal induced propagators survives depends on the loop integrations. The factors of $\omega$ that such propagators carry make the contribution from a region of loop momentum vanish only if the remaining integral has a finite limit. We already saw this mechanism in the scalar contribution to the gauge kernel in appendix~\ref{app:component-kernels}. The $h$ bubble is singular when either of the internal momenta becomes small. Near each region, the radial integral scales as $\int_0^\Lambda d\ell\,\ell^{2\omega-1}$, producing a $1/\omega$ pole. In more complicated graphs, such soft regions of integration can scale as $(\omega-\varepsilon)^{-1}$, as in~\eqref{eq:brst-ddr-soft}. In DDR, taking $D\to d$ first turns this into a $1/\omega$ pole, which can compensate an induced propagator's zero. Under our prescribed order of limits~\eqref{eq:brst-ddr-limits}, such a cancellation can leave a finite contribution from the soft region, even when the integrand vanishes at generic loop momentum. Higher order endpoint singularities can in turn lead to stronger enhancements.

Analogously, terms in interaction vertices that vanish at fixed momenta can similarly contribute after integration. As we discussed in the previous section, we can use the simpler logarithmic rule \eqref{eq:brst-Aphiphi-2d} when the transverse evanescent terms vanish after loop integrations. In general this is not allowed and we may need to use the complete
finite-$d$ interaction vertices, like the one given in~\eqref{eq:brst-Aphiphi-d}. In practice, it is safest to define the induced kernels and vertices at $d$, and integrate the loops in $D<d$, taking the limit $D\to d$ first and only then $d\to2$. This gives a BRST-consistent prescription for local correlation functions and anomalous dimensions.

It is natural to interpret the sensitivity to arbitrarily soft momenta as a momentum-space manifestation of the non-compact
throat in the small-instanton target space. We should be careful however, since the perturbative
prescription does not necessarily determine the nonperturbative
infrared state space, the treatment of global zero modes or the
vacuum central charge. These remain questions about the infrared
completion of the induced description.

\section{Outlook}
\label{sec:brst-consistency}

We have formulated a perturbative BRST quantisation of Witten's
induced gauge theory near the Higgs-branch singularity. The nonlocal
gauge fixing gives the transverse, longitudinal and ghost kernels
uniform scaling, while DDR specifies the regularization. The induced quadratic kernels and the leading $A\phi^2$
Slavnov--Taylor identity provide explicit checks of this formulation.
Following Witten~\cite{Witten:1997yu}, we expect the induced theory
to be conformal and refer to it as an $\mathrm{iCFT}_2$. Establishing
its conformal Ward identities within this prescription remains an
important next step. The supersymmetry of the induced action will
be examined in the companion paper~\cite{Stefanski:companion}.

The colour expansion allows planar contractions to be selected
before loop integration. A concrete application is to extend the
single-trace anomalous-dimension calculation
of~\cite{OhlssonSax:2014jtq} and test the weak-coupling spectrum
suggested by the QSC~\cite{Ekhammar:2026ykk}. This requires controlling
operator mixing and the ordered limits in diagrams with induced
internal lines.

The relation of these local operators to the unprotected discrete spectrum and
the throat continuum~\cite{Seiberg:1999xz,Aharony:1999dw}, including the role of
adjoint-scalar zero modes, also deserves study. More broadly, it
would be valuable to understand how this description relates to
other tractable regimes of $\AdS_3/\CFT_2$, including the distinct
pure NS-NS $k=1$ theory and its symmetric-orbifold
dual~\cite{Eberhardt:k1spectrum}. Finally, the induced-gauge theory construction could be generalised to other charged-matter theories and it would be worthwhile to explore these and determine under what conditions, such as lack on anomalies, these give rise to $\mathrm{iCFT}_2$s.

\section*{Acknowledgments}

I am indebted to  Olof Ohlsson-Sax for sharing his insights during our many conversations about the induced gauge theory description of the Higgs-branch CFT${}_2$. I would like to thank A.~Cavaglià, S.~Ekhammar, N.~Gromov,  C.~Thull and A.~Torrielli for recent related collaborations and many valuable discussions, to C.~Hull, A.~Tseytlin and K.~Zarembo for conversations and comments on the manuscript. I would also like to thank N.~Bobev, S.~Chester, J.~van~Muiden, S.~Pufu, R.~Reid-Edwards, and X.~Yin for stimulating exchanges. I acknowledge funding support from an STFC Consolidated Grant ‘Theoretical Particle Physics at City, University of London' ST/T000716/1. I am grateful to Anthropic and OpenAI for access to their models through the Claude Team plan for scientists and the ChatGPT for Academic Researchers programme. These have been used to test ideas, perform and check computations and help with the editing of the manuscript.

\appendix

\section{Lagrangians and field conventions}
\label{app:microscopic}

In this appendix we summarise the Lagrangians and field conventions used in this paper. We largely follow the Lorentzian component actions and conventions of~\cite{OhlssonSax:2014jtq},\footnote{We fix several typos found in~\cite{OhlssonSax:2014jtq}.  The $\phi$ kinetic term in $\mathcal{L}_V$ is gauge-covariant. When using a bispinor, rather than vector, index  for $\phi$ $\phi^{\alpha\dot\alpha}=\phi_i(\sigma^i)^{\alpha\dot\alpha}$, a factor of 2 appears since
$\phi_{\alpha\dot\alpha}\phi^{\alpha\dot\alpha}=2\phi_i\phi_i$, that had been missed. Because of this, we also redefine $T_{\mbox{here}}=\sqrt{2}T_{\mbox{there}}$ so that the kinetic terms for $T$ and $\phi$ are normalized in the same way. The sign of the $[\phi,\phi]^2$ and $[\phi,T]^2$ terms is now consistent with the positivity of the potential. 
The Yukawa couplings on the last lines of~\eqref{eq:uv-fund} and~\eqref{eq:uv-adj} have been bracketed into hermitian-conjugate pairs following our reality conventions, \textit{cf.}~\eqref{eq:H-psi-lambda-Yukawa-signs}. The \textit{overall} signs of these terms are not fixed and can be switched by changing the sign of $H$ and $T$. The \textit{relative} sign between the bracketed pairs is fixed in each case by observing that these terms all come from Kaluza-Klein reductions from six-dimensions and that the indices $\alpha$ and $\dot\alpha$  have opposite $\Gamma^{6789}$ chiralities. The $\sqrt{2}$  factors the $\mathcal L_H$ terms are needed to ensure that a supersymmetric $H$ vev gives equal masses to the bosonic and fermionic vector multiplet fields in the conventional UV setting. In $\mathcal{L}_T$ the two Yukawa couplings to $\phi$ pair \textit{opposite} chiralities of $\chi$. None of these changes affect the anomalous dimension calculations in~\cite{OhlssonSax:2014jtq}, since the only non-zero part comes from $\lambda$ fermion boxes and $\phi$ external states and the $\lambda^\dagger\phi\lambda$ vertices remain unchanged.} but continue to Euclidean signature as described below. We also summarising the Feynman rules for the fundamental hyper-multiplet fields that follow form the Lagrangains below.

The UV vector multiplet Lagrangian is
\begin{equation}\label{eq:uv-vector}
\begin{aligned}
\mathcal L_V=\tr\Bigl(&
-\tfrac14 F_{\mu\nu}F^{\mu\nu}
-\tfrac14\nabla_\mu\phi^{\alpha\dot\alpha}\nabla^\mu\phi_{\alpha\dot\alpha}
+i\psi^\dagger_{\sL\,\alpha\dot a}\nabla_+\psi_{\sL}^{\alpha\dot a}
+i\psi^\dagger_{\sR\,\dot\alpha\dot a}\nabla_-\psi_{\sR}^{\dot\alpha\dot a}
\\&
+\tfrac1{16}\bigl[\phi_{\alpha\dot\alpha},\phi_{\beta\dot\beta}\bigr]
\bigl[\phi^{\alpha\dot\alpha},\phi^{\beta\dot\beta}\bigr]
+i\psi^\dagger_{\sL\,\alpha\dot a}
\bigl[\phi^{\alpha\dot\alpha},\psi_{\sR\,\dot\alpha}{}^{\dot a}\bigr]
+i\psi^\dagger_{\sR\,\dot\alpha\dot a}
 \bigl[\phi^{\alpha\dot\alpha},\psi_{\sL\,\alpha}{}^{\dot a}\bigr]
+\tfrac12 D_{\dot a\dot b}D^{\dot a\dot b}\Bigr)\,.
\end{aligned}
\end{equation}
The fundamental hypermultiplet action is
\begin{equation}\label{eq:uv-fund}
\begin{aligned}
\mathcal L_H={}&
-\tfrac12\nabla_\mu H^\dagger_{\dot a}\nabla^\mu H^{\dot a}
+i\lambda^\dagger_{\sL\,\dot\alpha}\nabla_+\lambda_{\sL}^{\dot\alpha}
+i\lambda^\dagger_{\sR\,\alpha}\nabla_-\lambda_{\sR}^{\alpha}
+\tfrac12 F^\dagger_{\dot a}F^{\dot a}
\\&
-\tfrac14 H^\dagger_{\dot a}\phi_{\alpha\dot\alpha}\phi^{\alpha\dot\alpha}H^{\dot a}
+i\lambda^\dagger_{\sL\,\dot\alpha}\phi^{\alpha\dot\alpha}\lambda_{\sR\,\alpha}
+i\lambda^\dagger_{\sR\,\alpha}\phi^{\alpha\dot\alpha}\lambda_{\sL\,\dot\alpha}
+\tfrac12 H^\dagger_{\dot a}D^{\dot a\dot b}H_{\dot b}
\\&
+i\sqrt{2}\left(H^\dagger_{\dot a}\psi_{\sL}^{\dagger\,\alpha\dot a}\lambda_{\sR\,\alpha}
+\lambda^\dagger_{\sR\,\alpha}\psi_{\sL}^{\alpha\dot a}H_{\dot a}\right)
-i\sqrt{2}\left(H^\dagger_{\dot a}\psi_{\sR}^{\dagger\,\dot\alpha\dot a}\lambda_{\sL\,\dot\alpha}
+\lambda^\dagger_{\sL\,\dot\alpha}\psi_{\sR}^{\dot\alpha\dot a}H_{\dot a}\right)\,,
\end{aligned}
\end{equation}
and that of the adjoint hypermultiplet is
\begin{equation}\label{eq:uv-adj}
\begin{aligned}
\mathcal L_T=\tr\Bigl(&
-\tfrac14\nabla_\mu T_{a\dot a}\nabla^\mu T^{a\dot a}
+i\chi^\dagger_{\sL\,\dot\alpha a}\nabla_+\chi_{\sL}^{\dot\alpha a}
+i\chi^\dagger_{\sR\,\alpha a}\nabla_-\chi_{\sR}^{\alpha a}
+\tfrac12 G_{a\dot a}G^{a\dot a}
+\tfrac1{8}\bigl[\phi_{\alpha\dot\alpha},T_{a\dot a}\bigr]
 \bigl[\phi^{\alpha\dot\alpha},T^{a\dot a}\bigr]
\\&
+i\chi^\dagger_{\sL\,\dot\alpha a}
 \bigl[\phi^{\alpha\dot\alpha},\chi_{\sR\,\alpha}{}^{a}\bigr]
+i\chi^\dagger_{\sR\,\alpha a}
\bigl[\phi^{\alpha\dot\alpha},\chi_{\sL\,\dot\alpha}{}^{a}\bigr]
+\tfrac14 T_{a\dot a}\bigl[D^{\dot a\dot b},T^a{}_{\dot b}\bigr]
\\&
+i\Bigl(
 \psi^\dagger_{\sR\,\dot\alpha\dot a}[T_a{}^{\dot a},\chi_{\sL}^{\dot\alpha a}]
+\chi^\dagger_{\sL\,\dot\alpha a}[T^a{}_{\dot a},\psi_{\sR}^{\dot\alpha\dot a}]
\Bigr)
-i\Bigl(
 \psi^\dagger_{\sL\,\alpha\dot a}[T_a{}^{\dot a},\chi_{\sR}^{\alpha a}]
+\chi^\dagger_{\sR\,\alpha a}[T^a{}_{\dot a},\psi_{\sL}^{\alpha\dot a}]
\Bigr)
\Bigr)\,.
\end{aligned}
\end{equation}
Our  conventions for all $SU(2)$ indices are: $\epsilon^{12}=\epsilon_{12}=1$,
$X^i=\epsilon^{ij}X_j$ and $X_i=X^j\epsilon_{ji}$.
Hermitian conjugation reverses order without an extra graded sign.
In a Hermitian colour basis, the adjoint fermion reality conditions are
\begin{equation}
\begin{aligned}
(\psi_{\sL}^{\alpha\dot a})^\dagger
 &=\psi^\dagger_{\sL\,\alpha\dot a}=-\psi_{\sL\,\alpha\dot a},&\qquad\qquad
(\psi_{\sR}^{\dot\alpha\dot a})^\dagger
 &=\psi^\dagger_{\sR\,\dot\alpha\dot a}=+\psi_{\sR\,\dot\alpha\dot a},\\
(\chi_{\sL}^{\dot\alpha a})^\dagger
 &=\chi^\dagger_{\sL\,\dot\alpha a}=-\chi_{\sL\,\dot\alpha a},&\qquad\qquad
(\chi_{\sR}^{\alpha a})^\dagger
 &=\chi^\dagger_{\sR\,\alpha a}=+\chi_{\sR\,\alpha a}.
\end{aligned}
\end{equation}
The fundamental fermions are complex and unconstrained, with
\begin{equation}
(\lambda_{\sL}^{\dot\alpha})^\dagger=\lambda^\dagger_{\sL\,\dot\alpha}\,,\qquad\qquad\mbox{and}\qquad\qquad
(\lambda_{\sR}^\alpha)^\dagger=\lambda^\dagger_{\sR\,\alpha}\,.
\end{equation}
Raised dagger fields mean epsilon-raised conjugates:
\begin{equation}
 \psi_{\sL}^{\dagger\alpha\dot a}
=\epsilon^{\alpha\beta}\epsilon^{\dot a \dot b}
\psi^\dagger_{\sL\beta\dot b}
\,,\qquad\qquad\qquad\qquad
\lambda^{\dagger \alpha}_{\sR}
=\epsilon^{\alpha\beta}\lambda^\dagger_{\sR\,\beta}\,,
\end{equation}
with same-sign analogous equations for $\psi_{\sR},\chi_{\sL},\chi_{\sR}$ and $\lambda_{\sL}$.
In particular, $\psi_{\sL}^{\dagger\alpha\dot a}=-\psi_{\sL}^{\alpha\dot a}$ and 
$(\lambda_{\sR\,\alpha})^\dagger=-\lambda^{\dagger \alpha}_{\sR}$.
Using $(H^{\dot a})^\dagger=H^\dagger_{\dot a}$ and $\epsilon^{\beta\alpha}\epsilon_{\alpha\gamma}=-\delta^\beta{}_\gamma$, we find
\begin{equation}\label{eq:H-psi-lambda-Yukawa-signs}
\begin{aligned}
\left(iH^\dagger_{\dot a}\psi_{\sL}^{\dagger\alpha\dot a} \lambda_{\sR\,\alpha}\right)^\dagger
&=\left(-iH^\dagger_{\dot a}\psi_{\sL}^{\alpha\dot a}
\lambda_{\sR\,\alpha}\right)^\dagger
=+i(\lambda_{\sR\,\alpha})^\dagger
      (\psi_{\sL}^{\alpha\dot a})^\dagger H^{\dot a}\\
&=-i\epsilon^{\beta\alpha}\lambda^\dagger_{\sR\,\beta}\epsilon_{\alpha\gamma}\epsilon_{\dot a\dot b}\psi_{\sL}^{\gamma\dot b}H^{\dot a}
=+i\lambda^\dagger_{\sR\,\gamma}\psi_{\sL}^{\gamma\dot b}H_{\dot b}\,.
\end{aligned}
\end{equation}

Rotating to Euclidean space, we set $t=-ix_2$, $x=x_1$ and 
\begin{equation}
 A_{{\rm M},0}=iA_{{\rm E},2},\qquad
 A_{{\rm M},1}=A_{{\rm E},1},\qquad
 \phi_{\rm M}=\phi_{\rm E},\quad
 H_{\rm M}=H_{\rm E},\quad T_{\rm M}=T_{\rm E}.
\end{equation}
Here ${\rm M}$ denotes the Lorentzian fields in the Lagrangians above. The continued action satisfies $iS_{\rm M}=-S_{\rm E}$, and so the Euclidean functional integral weight is $e^{-S_{\rm E}}$. We write $p_\pm=p_1\pm ip_2$ and $A_\pm=A_1\pm iA_2$, with $p^2=p_+p_-$ in the physical two-plane.
We use the Fourier convention $e^{ip\cdot x}$.
The  derivatives are $\nabla_{{\rm M},\pm}=\nabla_t\pm\nabla_x$ and $\nabla_{{\rm E},\pm}=\nabla_1\pm i\nabla_2$ and the rotation gives
\begin{equation}\label{eq:brst-derivative-continuation}
 \nabla_{{\rm M},+}\longmapsto\nabla_{{\rm E},+},\qquad
 \nabla_{{\rm M},-}\longmapsto-\nabla_{{\rm E},-}.
\end{equation}
For the fundamental and vector fermions we choose the continued spinor
frame so that only phases enter the field identification. All fields
on the right-hand sides are Euclidean:
\begin{equation}\label{eq:brst-euclidean-fields}
\begin{aligned}
 \lambda_{{\rm M},\sL}&=-i\lambda_{\sL}\,,\qquad&
 \lambda^\dagger_{{\rm M},\sL}&=-\bar\lambda_{\sL}\,,\qquad &
 \lambda_{{\rm M},\sR}&=\lambda_{\sR}\,,\qquad &
 \lambda^\dagger_{{\rm M},\sR}&=-i\bar\lambda_{\sR},\\
 \psi_{{\rm M},\sL}&=\psi_{\sL}\,,\qquad &
 \psi^\dagger_{{\rm M},\sL}&=i\bar\psi_{\sL}\,,\qquad &
 \psi_{{\rm M},\sR}&=i\psi_{\sR}\,,\qquad &
 \psi^\dagger_{{\rm M},\sR}&=-\bar\psi_{\sR}\,.
\end{aligned}
\end{equation}
Above, index positions are unchanged and the barred variables are analytically continued
Grassmann partners, rather than ordinary Euclidean conjugates.
The adjoint fermion and auxiliary field are continued as
\begin{equation}
 \chi^\dagger\longmapsto-i\bar\chi_{\rm E}\,,\qquad\qquad
 D_{\rm M}=iD_{\rm E}\,.
\end{equation}
The hypermultiplet auxiliaries $F$ and $G$ can be eliminated algebraically before continuing. Throughout the paper we drop the subscript ${\rm E}$, writing $\phi,\psi,H,\lambda$, etc for the Euclidean fields. 

It is useful to combine the  fundamental fermions defining $\Lambda=(\lambda_{\sL},\lambda_{\sR})^{\mathsf T}$ and
$\bar\Lambda=(\bar\lambda_{\sR},\bar\lambda_{\sL})$. With
$\gamma^1=\sigma_1$, $\gamma^2=\sigma_2$, 
their kinetic and scalar terms are
\begin{equation}\label{eq:brst-euclidean-dirac-operator}
 \mathcal L_{{\rm E},\lambda}
 =\bar\Lambda
\left(\gamma^\mu\nabla_\mu\otimes\mathbf1_2+\mathcal M(\phi)\right)\Lambda\,,
\end{equation}
where $\mathcal M$, with column indices $(\dot\beta,\beta)$ and row indices $(\alpha,\dot\alpha)$ is
\begin{equation}\label{eq:brst-dirac-scalar-matrix}
\begin{aligned}
 \mathcal M(\phi)&=-i
 \begin{pmatrix}
\phi^{\alpha\dot\gamma}\epsilon_{\dot\gamma\dot\beta}&0\\
 0&\phi^{\gamma\dot\alpha}\epsilon_{\gamma\beta}
 \end{pmatrix}.
\end{aligned}
\end{equation}
Differentiating with respect to the bispinor gives the constant matrix\footnote{Explicitly,
$(\phi^{\alpha\dot\alpha})=-i\phi_1\sigma_1+i\phi_2\sigma_2
+\phi_3\mathbf1_2+i\phi_4\sigma_3$, with $\phi_I$ real. For a real scalar leg $\phi_I$, the vertex is
$V_{\phi_I}^{a}=\mathcal M_I T^a$, where
$\mathcal M_I=(\partial\phi^{\gamma\dot\delta}/\partial\phi_I)
\mathcal M_{\gamma\dot\delta}$ is the same constant vertex in that basis.}
\begin{equation}\label{eq:brst-bispinor-scalar-vertex}
 \mathcal M_{\gamma\dot\delta}
 \equiv\frac{\partial\mathcal M}{\partial\phi^{\gamma\dot\delta}}
 =-i\begin{pmatrix}
\delta^\alpha{}_{\gamma}\epsilon_{\dot\delta\dot\beta}&0\\
 0&\delta^{\dot\alpha}{}_{\dot\delta}\epsilon_{\gamma\beta}
 \end{pmatrix}.
\end{equation}

The Euclidean~\eqref{eq:brst-euclidean-fields}
propagators and gauge vertices involving the fundamental fields follow from kinetic terms in~\eqref{eq:uv-fund} and are 
\begin{equation}\label{eq:brst-fund-propagator-residues}
 \langle H_{\dot a}(\ell)H^\dagger_{\dot b}(-\ell)\rangle_0
 =\frac{2\delta_{\dot a\dot b}}{\ell^2}\,,
 \quad
  \langle\lambda_{\sL}(\ell)\bar\lambda_{\sL}(-\ell)\rangle_0 
 =-i\frac{\ell_-}{\ell^2}\,,
 \quad
  \langle\lambda_{\sR}(\ell)\bar\lambda_{\sR}(-\ell)\rangle_0=-i\frac{\ell_+}{\ell^2}\,,
\end{equation}
where $\ell^2=\ell_+\ell_-$ in the physical two-plane; in the continued loop integrals, $\ell^2=\ell_+\ell_-+\hat\ell^{\,2}$. Expanding
\(\tfrac12|\nabla H|^2\), the scalar gauge vertices are
\begin{equation}\label{eq:brst-H-gauge-rules}
 V^{a}_{AHH,\mu}(\ell+p,\ell)
 =\frac{i}{2}(2\ell+p)_\mu T^a,
 \qquad
 V^{ab}_{AAHH,\mu\nu}
 =\frac12\delta_{\mu\nu}\{T^a,T^b\}.
\end{equation}
We see that the factor 2 in each $H$ propagator cancels the factor $1/2$ at each trilinear gauge vertex.  We keep
\(\operatorname{tr}(T^aT^b)=T_F\delta^{ab}\) throughout.

\section{BRST conventions and gauge choices}
\label{app:brst}

\subsection{Nilpotency of BRST transformations}

The BRST transformations given in section~\ref{sec:brst-fixing} are nilpotent. $c,\bar c$ are Grassmann odd and $b$ Grassmann even. We write
$\delta\equiv\delta_{\mathrm{BRST}}$, $Q\equiv Q_{\mathrm{BRST}}$ with the BRST variation as graded commutator with the odd, ghost-number-one charge $Q$,
\begin{equation}\label{eq:brst-Q}
\delta X=[Q,X\}\equiv QX-(-1)^{|X|}XQ \,.
\end{equation}
It commutes with $\partial_{\mu}$ and obeys the \emph{left} Leibniz rule
\begin{equation}\label{eq:leibniz}
\delta(XY)=(\delta X)Y+(-1)^{|X|}X(\delta Y).
\end{equation}
The transformations are\footnote{\label{ftnt:hermiticity} With $c$ Hermitian one has $(c^{2})^{\dagger}=c^{2}$. So $\delta c=ic^{2}$ is \emph{anti}-Hermitian even though $c$ is Hermitian, and
$\delta\bar c=b$ then forces $b^{\dagger}=-b$. }
\begin{equation}\label{eq:brst-app}
\delta A_{\mu}=\nabla_{\mu}c,\qquad
\delta c=ic^{2},\qquad
\delta\bar c=b,\qquad
\delta b=0,\qquad
\delta\phi=i[c,\phi],\qquad
\delta H=icH \,.
\end{equation}
For either parity of $X$ applying \eqref{eq:brst-Q} twice we find
$\delta^{2}X=[Q^{2},X]$ with $Q^{2}=\tfrac12\{Q,Q\}$, so $Q^{2}=0$ is equivalent to
$\delta^{2}=0$ on each field. To verify this we note the identity
\begin{equation}\label{eq:lemma}
\{Y,[Y,X]\}=[Y^{2},X],
\end{equation}
valid for arbitrary matrices $X$ and $Y$.

\noindent \textbf{Ghost.} Since $|c|=1$, \eqref{eq:leibniz} gives
$\delta(c^{2})=(\delta c)c-c(\delta c)=[\delta c,c]$, and therefore
\begin{equation}
\delta^{2}c=i\,[\,ic^{2},c\,]=i^{2}\big(c^{3}-c^{3}\big)=0\,.
\end{equation}
\noindent \textbf{Fundamental matter.} Leaving $\delta c$ undetermined,
\begin{equation}
\delta^{2}H=\delta(icH)=i(\delta c)H-ic\,(\delta H)
=i(\delta c)H-ic\,(icH)=i(\delta c)H+c^{2}H \,.
\end{equation}
This vanishes for generic $H$ when $\delta c=ic^{2}$. 

\noindent \textbf{Adjoint scalar.} From $\delta(c\phi)=(\delta c)\phi-c\,\delta\phi$ and
$\delta(\phi c)=(\delta\phi)c+\phi\,\delta c$ one gets
$\delta[c,\phi]=[\delta c,\phi]-\{c,\delta\phi\}$, so
\begin{equation}
\delta^{2}\phi=i\Big(i[c^{2},\phi]-i\{c,[c,\phi]\}\Big)
=-\Big([c^{2},\phi]-\{c,[c,\phi]\}\Big)=0
\end{equation}
by \eqref{eq:lemma} with $Y=c$, $X=\phi$.

\noindent \textbf{Gauge field.} Similarly
$\delta[A_{\mu},c]=\{\delta A_{\mu},c\}+[A_{\mu},\delta c]$, the anticommutator
arising because $\delta A_{\mu}$ is odd. Using
$\partial_{\mu}(c^{2})=\{\partial_{\mu}c,c\}$,
\begin{align}
\delta^{2}A_{\mu}
&=\partial_{\mu}(\delta c)-i\{\nabla_{\mu}c,c\}-i[A_{\mu},\delta c]\nonumber\\
&=i\{\partial_{\mu}c,c\}-i\{\partial_{\mu}c,c\}-\{[A_{\mu},c],c\}+[A_{\mu},c^{2}]
=0 ,
\end{align}
since $\{[A_{\mu},c],c\}=-\{c,[c,A_{\mu}]\}=-[c^{2},A_{\mu}]=[A_{\mu},c^{2}]$, again
by \eqref{eq:lemma}. 

\noindent \textbf{Trivial pair.} Finally $\delta^{2}\bar c=\delta b=0$ and $\delta^{2}b=0$ by
construction.

\subsection{Equivalent representatives and other gauges}
\label{sec:brst-gauge-representatives}

The dimensions of the pair $(\bar c,b)$ in
\eqref{eq:brst-dims} depend on our choice of variables. For any real number
$r$ for which $K^r$ is defined on the chosen nonzero-mode space, set
\begin{equation}\label{eq:brst-r-redefinition}
\bar c_r=K^r\bar c,\qquad b_r=K^r b\,.
\end{equation}
Because $K$ is field-independent, we still have
$\delta\bar c_r=b_r$ and $\delta b_r=0$ however their dimensions now are $\dim\bar c_r=\dim b_r=2+r(d-4)$, or $2-2r$ at $d=2$.
With a common finite-mode regulator, the bosonic Jacobian
$(\det K^r)^{-1}$ cancels the Berezin Jacobian $\det K^r$.
Transforming sources along with the fields therefore leaves the
gauge-invariant correlators, Slavnov--Taylor identities and BRST cohomology unchanged. In these variables \eqref{eq:brst-sPsi} becomes
\begin{equation}\label{eq:brst-sPsi-r}
\delta\Psi=\zeta\Nf\,\tr\!\int d^dx\left[
i\,b_rK^{1-r}\partial\!\cdot\!A
+\frac{\xi}{2}b_rK^{1-2r}b_r
-i\,\bar c_rK^{1-r}\partial\!\cdot\!\nabla c
\right]\,.
\end{equation}
Eliminating $b_r$ gives 
\eqref{eq:brst-gf} as before for all $r$, so the longitudinal gauge kernel is unchanged.

\noindent\textbf{Local Faddeev-Popov ghosts.}
For $r=1$, define $\bar c_0\equiv K\bar c$ and $b_0\equiv Kb$. The action becomes
\begin{equation}\label{eq:brst-sPsi-zero}
\delta\Psi=\zeta\Nf\,\tr\!\int d^dx\left[
i\,b_0\,\partial\!\cdot\!A
+\frac{\xi}{2}b_0K^{-1}b_0
-i\,\bar c_0\,\partial\!\cdot\!\nabla c
\right].
\end{equation}
The Faddeev-Popov operator is local and the nonlocal kernel now appears
only in the Gaussian weight for the gauge condition. Here
$\dim c=0$ and $\dim\bar c_0=\dim b_0=d-2$, so all three fields have dimension zero at $d=2$.

\noindent\textbf{Landau gauge.}
Another familiar gauge choice can be made by setting $\xi=0$ in \eqref{eq:brst-sPsi-zero}. It gives a local gauge fermion and Faddeev-Popov action, with
the $b_0$ integral imposing $\partial\!\cdot\!A=0$. The
gauge propagator is the purely transverse $\xi\to0$ limit of \eqref{eq:brst-AA}.

\noindent\textbf{An ordinary local covariant gauge.}
Another choice is
\begin{equation}\label{eq:brst-local-Psi}
\Psi_{\rm loc}
=\zeta\Nf\,\tr\!\int d^dx\,
\bar c_0\left(i\,\partial\!\cdot\!A+\frac{\alpha}{2}b_0\right),
\qquad [\alpha]=4-d \,.
\end{equation}
Here the ghosts are local and the gauge-fixing term is
$(\partial\!\cdot\!A)^2/(2\alpha)$. The longitudinal inverse
propagator then scales as $p^2/\alpha$, whereas the induced transverse
kernel scales as $(p^2)^\omega$. The gauge inherited from the UV
action is of this type, with $\alpha$ proportional to $g_{\rm YM}^2$
and vanishing in the IR limit. Section~\ref{sec:brst-fixing} instead keeps $\xi$ dimensionless without adding a scale, and leads to
ghost  and gauge propagators with the same momentum dependence and $1/\Nf$ scaling.

\section{Master integrals}
\label{app:masters}
The DDR integrals needed for the induced diagrams in
\eqref{eq:inducedblobs} include 
\begin{equation}
\intq{q}\frac{q_{-}}{q^{2}(q+p)^{2}}=-\tfrac{p_{-}}{2}B_{d,D},
\qquad
\intq{q}\frac{q_{-}(q+p)_{-}}{q^{2}(q+p)^{2}}
=-\tfrac{D-2}{4(D-1)}\,p_{-}^{2}B_{d,D}\,.
\label{eq:moments}
\end{equation}
These follow from an ordinary Feynman-parameter calculation.\footnote{The measure factor
$\mu^{d-D}=\mu^{\ereg}$ must be retained through pole expansion and
ultraviolet subtraction at fixed $\wreg$: its expansion against a
$1/\ereg$ pole contributes finite scale logarithms.}
We collect this class of integrals in the master formula
\begin{equation}
\begin{aligned}
J(m,n;a,b)
&=\intq{q}\frac{q_{+}^{m}\,(\hat q^{\,2})^{n}}{(q^{2})^{a}
\,((q+p)^{2})^{b}}
\\
&=
{\mu^{d-D}}\,
\frac{(-p_{+})^{\,m}\,(p^{2})^{\frac D2+n-a-b}}{(4\pi)^{D/2}}\,
\frac{\Gamma(\tfrac D2-1+n)}{\Gamma(\tfrac D2-1)}
\\
&\qquad{}\times
\frac{\Gamma(a{+}b{-}n{-}\tfrac D2)\,
\Gamma(\tfrac D2{+}n{+}m{-}a)\,\Gamma(\tfrac D2{+}n{-}b)}
{\Gamma(a)\,\Gamma(b)\,\Gamma(D{+}2n{+}m{-}a{-}b)},
\end{aligned}
\label{eq:Jmaster}
\end{equation}
for integers $m,n\ge0$ and external momentum in the physical two-plane,
$\hat p=0$. We derive this formula below and then explain the vanishing of
the scaleless tadpoles.

\subsection{Derivation of the master integral}

We work in $D$ Euclidean dimensions, with the transverse $(D-2)$ vector
denoted by $\hat q$. The restriction $\hat p=0$ is a kinematic assumption
of the formula; it does not follow from covariance.

Combining the two denominators in \eqref{eq:Jmaster} gives
\begin{equation}
  \frac{1}{(q^{2})^{a}\,\bigl((q+p)^{2}\bigr)^{b}}
  =\frac{\Gamma(a+b)}{\Gamma(a)\Gamma(b)}
  \int_{0}^{1}\!dx\;
  \frac{x^{\,b-1}(1-x)^{\,a-1}}
       {\bigl[(1-x)q^{2}+x(q+p)^{2}\bigr]^{a+b}} ,
\end{equation}
where the denominator can be written as
\begin{equation}
  (1-x)q^{2}+x(q+p)^{2}=(q+xp)^{2}+\Delta,
  \qquad \Delta=x(1-x)\,p^{2}.
\end{equation}
Shifting $q\to q-xp$ (with $\hat p=0$, so $\hat q$ is untouched) gives
\begin{equation}
  J={\mu^{d-D}}\,\frac{\Gamma(a+b)}{\Gamma(a)\Gamma(b)}
  \int_{0}^{1}\!dx\;x^{\,b-1}(1-x)^{\,a-1}
  \int\!\frac{d^{D}q}{(2\pi)^{D}}\,
  \frac{(q_{+}-x\,p_{+})^{m}\,(\hat q^{\,2})^{n}}
       {(q^{2}+\Delta)^{a+b}} .
\end{equation}
We split the loop momentum into the two longitudinal directions
$q_{1},q_{2}$ and the $(D-2)$ transverse directions $\hat q$, so that
$d^{D}q=dq_{1}\,dq_{2}\,d^{D-2}\hat q$, and use the Schwinger parametrisation
\begin{equation}
  \frac{1}{(q^{2}+\Delta)^{a+b}}
  =\frac{1}{\Gamma(a+b)}\int_{0}^{\infty}\!dt\;t^{a+b-1}\,
   e^{-t\,(q^{2}+\Delta)} .
\end{equation}
In the expansion
$(q_{+}-x p_{+})^{m}=\sum_{j}\binom{m}{j}q_{+}^{\,j}(-x p_{+})^{m-j}$,
only the $j=0$ term survives the momentum integrations, since $q_+$ is
null. The Gaussian momentum integrals leave a Schwinger integral of the form
\begin{equation}
  \int_{0}^{\infty}\!dt\;
  t^{a+b-n-\frac D2-1}\,e^{-t\Delta}
  =\Gamma\!\Bigl(a+b-n-\tfrac D2\Bigr)\,
   \Delta^{\frac D2+n-a-b} .
\end{equation}
Together these give
\begin{equation}
  \int\!\frac{d^{D}q}{(2\pi)^{D}}\,
  \frac{(\hat q^{\,2})^{n}}
       {(q^{2}+\Delta)^{a+b}} 
  =\frac{1}{(4\pi)^{D/2}}
   \frac{\Gamma\!\bigl(n+\tfrac{D-2}{2}\bigr)\,
   \Gamma\!\bigl(a+b-n-\tfrac D2\bigr)}{\Gamma(\tfrac{D-2}{2})\Gamma(a+b)}\;
   \,
   \Delta^{\frac D2+n-a-b} .
\end{equation}
The remaining $x$ integral is
\begin{equation}
  \int_{0}^{1}\!dx\;x^{\,\tfrac D2+m+n-a-1}(1-x)^{\,\tfrac D2+n-b-1}
  =\frac{\Gamma\!\bigl(\tfrac D2+n+m-a\bigr)
  \,\Gamma\!\bigl(\tfrac D2+n-b\bigr)}{\Gamma(D+2n+m-a-b)}.
\end{equation}
Combining these factors gives \eqref{eq:Jmaster}.

\subsection{Scaleless tadpoles}
\label{app:tadpoles}
The tadpole diagrams in~\eqref{eq:inducedblobs} come from quartic vertices in~\eqref{eq:uv-fund}. Each involves an $H$ line closing on itself which gives a scaleless loop integral
\begin{equation}
T_{\alpha}=\intq{q}\,(q^{2})^{-\alpha},
\label{eq:tadpole}
\end{equation}
where $\alpha=1$. After the angular integration, we split the radial integral at an arbitrary scale $\mu$,
\begin{equation}
T_{\alpha}=\frac{\mu^{d-D}\Omega_D}{(2\pi)^D}
\left[
\int_0^\mu d|q|\,|q|^{D-2\alpha-1}
+\int_\mu^\infty d|q|\,|q|^{D-2\alpha-1}
\right],
\end{equation}
where $\Omega_D$ is the volume of $\Sphere^{D-1}$. The first term converges for $D>2\alpha$ and gives $+\frac{\Omega_D\,\mu^{d-2\alpha}}{(2\pi)^D(D-2\alpha)}$. We analytically continue it to all $D<2\alpha$. The second term converges for $D<2\alpha$ and gives $-\frac{\Omega_D\,\mu^{d-2\alpha}}{(2\pi)^D(D-2\alpha)}$, which we continue to all $D>2\alpha$. The two continued expressions cancel identically. At $D=2\alpha$, we first combine the two continued expressions and then take the limit. It is in this sense that $T_{\alpha}=0$ for all $D$ and all $\alpha$.

\section{The induced quadratic kernels}
\label{app:brst-quadratic}
\label{app:component-kernels}

In this appendix we compute the quadratic action and propagators summarised in section~\ref{sec:brst-rules}. We begin by collecting the
momentum integrals that enter the subsequent 1-loop using the master integral formula from appendix~\ref{app:masters}, and then proceed to compute the diagrams, including  the numerator algebra and
multiplicities for each field. We also explain how the reality conditions enter when the fermion kernel is inverted.

\paragraph{One-loop integrals.}
We will denote the dimension of the defining fundamental loop by $\widehat D$. All the
quadratic kernels we encounter can be expressed in terms of 
\begin{equation}\label{eq:brst-BdDhat}
\begin{aligned}
 B_{d,\widehat D}(p;\mu)
 &\equiv
 \mu^{d-\widehat D}
 \int\!\frac{d^{\widehat D}\ell}{(2\pi)^{\widehat D}}\,
 \frac{1}{\ell^2(\ell+p)^2}
\\
 &=
 \mu^{d-\widehat D}(4\pi)^{-\widehat D/2}
 \frac{\Gamma(2-\widehat D/2)
 \Gamma(\widehat D/2-1)^2}{\Gamma(\widehat D-2)}
 (p^2)^{\widehat D/2-2}\,.
\end{aligned}
\end{equation}
This scalar bubble converges for $2<\widehat D<4$ and is the $J(0,0;1,1)$ in \eqref{eq:Jmaster}. We will also encounter the following integrals 
\begin{subequations}\label{eq:brst-bubble-reductions}
\begin{equation}\label{eq:brst-bubble-first}
 \mu^{d-\widehat D}\int\!\frac{d^{\widehat D}\ell}{(2\pi)^{\widehat D}}
 \frac{\ell_\mu}{\ell^2(\ell+p)^2}
 =-\frac{p_\mu}{2}B_{d,\widehat D}(p),
\end{equation}
\begin{equation}\label{eq:brst-bubble-dot}
 \mu^{d-\widehat D}\int\!\frac{d^{\widehat D}\ell}{(2\pi)^{\widehat D}}
 \frac{\ell\cdot(\ell+p)}{\ell^2(\ell+p)^2}
 =-\frac{p^2}{2}B_{d,\widehat D}(p),
\end{equation}
\begin{equation}\label{eq:brst-bubble-null}
 \mu^{d-\widehat D}\int\!\frac{d^{\widehat D}\ell}{(2\pi)^{\widehat D}}
 \frac{\ell_\pm(\ell+p)_\pm}{\ell^2(\ell+p)^2}
 =-\frac{\widehat D-2}{4(\widehat D-1)}
 p_\pm^2B_{d,\widehat D}(p),
\end{equation}
\begin{equation}\label{eq:brst-bubble-tensor}
\begin{aligned}
 \mu^{d-\widehat D}\int\!\frac{d^{\widehat D}\ell}{(2\pi)^{\widehat D}}
 \frac{\ell_\mu(\ell+p)_\nu}{\ell^2(\ell+p)^2}
 &=-\frac{p^2\delta_{\mu\nu}+(\widehat D-2)p_\mu p_\nu}
 {4(\widehat D-1)}B_{d,\widehat D}(p),
\end{aligned}
\end{equation}
\begin{equation}\label{eq:brst-tadpole-zero}
 T_1\equiv
 \mu^{d-\widehat D}
 \int\!\frac{d^{\widehat D}\ell}{(2\pi)^{\widehat D}}\,
 \frac1{\ell^2}
 =\mu^{d-\widehat D}
 \int\!\frac{d^{\widehat D}\ell}{(2\pi)^{\widehat D}}\,
 \frac1{(\ell+p)^2}=0,
\end{equation}
\begin{equation}\label{eq:brst-vector-tadpole-zero}
 \mu^{d-\widehat D}
 \int\!\frac{d^{\widehat D}\ell}{(2\pi)^{\widehat D}}\,
 \frac{\ell_\mu}{\ell^2}
 =\mu^{d-\widehat D}
 \int\!\frac{d^{\widehat D}\ell}{(2\pi)^{\widehat D}}\,
 \frac{(\ell+p)_\mu}{(\ell+p)^2}=0.
\end{equation}
\end{subequations}

Equation \eqref{eq:brst-bubble-first} is $J(1,0;1,1)$ in \eqref{eq:Jmaster} when the numerator is $\ell_+$. Covariantising then gives the result for $\ell_\mu$. If the numerator is
instead $(\ell+p)_\mu$, we add $p_\mu B_{d,\widehat D}$
to obtain $+p_\mu B_{d,\widehat D}/2$.
For \eqref{eq:brst-bubble-dot}, we write
$2\ell\cdot(\ell+p)=\ell^2+(\ell+p)^2-p^2$. The first two terms
cancel one of the  denominators and become zero using~\eqref{eq:brst-tadpole-zero}. The remaining integral is gives $-p^2B_{d,\widehat D}/2$. 
Equation~\eqref{eq:brst-bubble-null}, follows from using~\eqref{eq:Jmaster} for $J(2,0;1,1)+p_\pm J(1,0;1,1)$. Note, that here, as in the master formula,
the external momentum lies in the physical two-plane.
For~\eqref{eq:brst-bubble-tensor}, we start with an integral whose numerator is $\ell_\mu\ell_\nu$. By covariance, the
integral must be a linear combination of $\delta_{\mu\nu}$ and $p_\mu p_\nu$. Its $++$ component follows from
\(J(2,0;1,1)\), while contracting its indices gives the vanishing tadpole $T_1$. These two conditions determine the coefficients of the $\ell_\mu\ell_\nu$ term.
Adding $p_\nu$ times \eqref{eq:brst-bubble-first} then gives
\eqref{eq:brst-bubble-tensor}. 

The tadpoles in \eqref{eq:brst-tadpole-zero} vanish by the argument
in appendix~\ref{app:tadpoles}. After shifting the integration variable,
the vector integrals in \eqref{eq:brst-vector-tadpole-zero} vanish
because their integrands are odd. Such shifts are allowed in
dimensional regularisation. 

\paragraph{Fundamental scalar loop contribution to gauge field two-point kernel.}
Let
\begin{equation}
 {\cal I}_{\mu\nu}(p)
 =\mu^{d-\widehat D}\int\!\frac{d^{\widehat D}\ell}{(2\pi)^{\widehat D}}
 \frac{(2\ell+p)_\mu(2\ell+p)_\nu}
 {\ell^2(\ell+p)^2}\,.
\end{equation}
The two complex components \(H^{\dot a}\), their propagator residues and
the vertices~\eqref{eq:brst-H-gauge-rules} give
\begin{equation}\label{eq:brst-A-H-diagrams}
 K^{ab,H}_{A,\mu\nu}(p)
 =2\Nf T_F\delta^{ab}
 \left[-{\cal I}_{\mu\nu}(p)+2\delta_{\mu\nu}T_1\right]\,.
\end{equation}
We have kept the bubble and seagull together, with their relative
coefficient explicit. The two scalar propagator residues cancel
the two factors of \(1/2\) at the trilinear vertices. The remaining
factor of two counts \(H^{\dot a}\), while the colour and flavour sums
give \(\Nf T_F\delta^{ab}\). Contracting the bubble with the external
momentum gives
\begin{equation}
 p^\mu{\cal I}_{\mu\nu}
 =2p_\nu T_1\,,
\end{equation}
which follows from
$p\cdot(2\ell+p)=(\ell+p)^2-\ell^2$ to cancel one denominator
in each term. The two remaining numerators become
\(2\ell_\nu+p_\nu\) and \(2\ell_\nu-p_\nu\) after shifting the
second integral. Their odd parts vanish by
\eqref{eq:brst-vector-tadpole-zero}, leaving \(2p_\nu T_1\).
Before using \(T_1=0\), the full scalar contribution obeys
\begin{equation}
 p^\mu K^{ab,H}_{A,\mu\nu}
 =2\Nf T_F\delta^{ab}\bigl[-2p_\nu T_1+2p_\nu T_1\bigr]=0,
\end{equation}
so the Ward identity holds before assigning a value to the scaleless
tadpole.
Taking the trace of the propagator, the numerator identity
$(2\ell+p)^2=2\ell^2+2(\ell+p)^2-p^2$ reduces the integral to
\eqref{eq:brst-BdDhat} and \eqref{eq:brst-tadpole-zero}:
\begin{equation}
 \delta^{\mu\nu}{\cal I}_{\mu\nu}
 =4T_1-p^2B_{d,\widehat D}(p)
 =-p^2B_{d,\widehat D}(p)\,.
\end{equation}
The tensor
\(p^2\delta_{\mu\nu}-p_\mu p_\nu\) has trace
\((\widehat D-1)p^2\), so its contribution is fixed.  Finally, including the sign in \eqref{eq:brst-A-H-diagrams} gives
\begin{equation}\label{eq:brst-A-H-result}
 K^{ab,H}_{A,\mu\nu}(p)
 =\frac{2\Nf T_F}{\widehat D-1}\delta^{ab}
 \bigl(p^2\delta_{\mu\nu}-p_\mu p_\nu\bigr)
 B_{d,\widehat D}(p).
\end{equation}
We see that we can set the seagull to zero without losing transversality in DDR.

\paragraph{Fundamental fermions loop contribution to  gauge field two-point kernel.}
The two \(\lambda_{\sL}\) and two \(\lambda_{\sR}\) components combine to form two
Dirac fermions per flavour. Combining the two chiralities before doing the
$\widehat D$-dimensional numerator algebra gives
\begin{equation}\label{eq:brst-A-lambda-start}
 K^{ab,\lambda}_{A,\mu\nu}(p)
 =2\Nf T_F\delta^{ab}{\cal F}_{\mu\nu}(p),
 \qquad
 {\cal F}_{\mu\nu}
 =\mu^{d-\widehat D}\int\!\frac{d^{\widehat D}\ell}{(2\pi)^{\widehat D}}
 \frac{\operatorname{tr}_\gamma[
 \gamma_\mu\slashed\ell\gamma_\nu(\slashed\ell+\slashed p)]}
 {\ell^2(\ell+p)^2}\,.
\end{equation}
with $\operatorname{tr}_\gamma\mathbf1=2$. The sign above comes from the the closed fermion loop and the two gauge vertices each contributing $-1$. The spinor trace in the numerator gives
\begin{equation}\label{eq:PiA-Ftrace}
\operatorname{tr}_\gamma[
\gamma_\mu\slashed\ell\gamma_\nu(\slashed\ell+\slashed p)]
=2\bigl[\ell_\mu(\ell+p)_\nu+\ell_\nu(\ell+p)_\mu
-\delta_{\mu\nu}\ell\cdot(\ell+p)\bigr]\,.
\end{equation}
For each term we can use~\eqref{eq:brst-bubble-dot} and \eqref{eq:brst-bubble-tensor} to perform the integrations. We can also extract the coefficient from its Ward identity and trace, as for
the scalar loop.
Contracting with \(p^\mu\) reduces
\({\cal F}_{\mu\nu}\) to a difference of scaleless tadpoles, so it is
transverse:
\begin{equation}\label{eq:brst-A-lambda-transverse}
\begin{aligned}
p^\mu{\cal F}_{\mu\nu}(p)
={}&
\mu^{d-\widehat D}
\int\!\frac{d^{\widehat D}\ell}{(2\pi)^{\widehat D}}\,
\frac{\operatorname{tr}_\gamma\!\left[
 \slashed p\,\slashed\ell\,\gamma_\nu
 (\slashed\ell+\slashed p)\right]}
{\ell^2(\ell+p)^2}
\\
={}&
\mu^{d-\widehat D}
\int\!\frac{d^{\widehat D}\ell}{(2\pi)^{\widehat D}}\,
\frac{
 \operatorname{tr}_\gamma\!\left[
 (\slashed\ell+\slashed p)\slashed\ell\gamma_\nu
 (\slashed\ell+\slashed p)\right]
 -
 \operatorname{tr}_\gamma\!\left[
 \slashed\ell\slashed\ell\gamma_\nu
 (\slashed\ell+\slashed p)\right]}
{\ell^2(\ell+p)^2}
\\
={}&
2\mu^{d-\widehat D}
\int\!\frac{d^{\widehat D}\ell}{(2\pi)^{\widehat D}}
\left[
 \frac{\ell_\nu}{\ell^2}
 -
 \frac{(\ell+p)_\nu}{(\ell+p)^2}
\right]
=0 ,
\end{aligned}
\end{equation}
The trace of ${\cal F}_{\mu\nu}$ is
\begin{equation}
 \delta^{\mu\nu}{\cal F}_{\mu\nu}
 =(2-\widehat D)
 \mu^{d-\widehat D}\int\!\frac{d^{\widehat D}\ell}{(2\pi)^{\widehat D}}
 \frac{\operatorname{tr}_\gamma[
 \slashed\ell(\slashed\ell+\slashed p)]}
 {\ell^2(\ell+p)^2}
=(\widehat D-2)p^2B_{d,\widehat D}(p).
\end{equation}
The last equality uses
\(\gamma^\mu\gamma^\rho\gamma_\mu=(2-\widehat D)\gamma^\rho\),
\(\operatorname{tr}_\gamma[\slashed\ell(\slashed\ell+\slashed p)]
=2\ell\cdot(\ell+p)\), and \eqref{eq:brst-bubble-dot}. As before,
dividing by the transverse trace \((\widehat D-1)p^2\) gives
\begin{equation}\label{eq:brst-A-lambda-result}
 K^{ab,\lambda}_{A,\mu\nu}(p)
 =\frac{2\Nf T_F(\widehat D-2)}{\widehat D-1}\delta^{ab}
 \bigl(p^2\delta_{\mu\nu}-p_\mu p_\nu\bigr)
 B_{d,\widehat D}(p).
\end{equation}
Adding \eqref{eq:brst-A-H-result} and
\eqref{eq:brst-A-lambda-result} gives
\begin{equation}\label{eq:brst-A-total-B}
 K^{ab}_{A,\mu\nu}(p)
 =2\Nf T_F\delta^{ab}
 \bigl(p^2\delta_{\mu\nu}-p_\mu p_\nu\bigr)
 B_{d,\widehat D}(p).
\end{equation}
The fermion diagrams are individually transverse, while the scalar
contribution is transverse after combining the bubble and seagull. In
both cases this follows from the calculation, without imposing a
transverse form on the answer.

The separate contributions also show where the pole at
$\widehat D=2$ occurs. The factor $\widehat D-2$ in
\eqref{eq:brst-A-lambda-result} cancels the bubble pole, leaving a finite
fermion contribution. No such factor multiplies
\eqref{eq:brst-A-H-result}, so the fundamental scalars carry the entire
pole of the gauge kernel. Its consequences for the induced propagators
are discussed in section~\ref{sec:evanescent}.

\paragraph{Regulating chiral two-point loops.}
\label{app:chiral-subloops}
The fundamental-fermion bubble in
\eqref{eq:brst-A-lambda-start} contributes to the gauge-field two-point
function. The left- and right-moving fermions $\lambda$ couple in a conventional way to the gauge field and so each have two-dimensional gauge anomalies. Since these cancel in our non-chiral theory, we combine them into one non-chiral Dirac fermion running in the loop. If we had chosen instead to calculate the chiral bubbles separately
in strictly two dimensions we would encounter a familiar problem.  $A_+$ couples only to $\lambda_{\sL}$, and $A_-$ only to $\lambda_{\sR}$ giving
\begin{equation}\label{eq:PiA-chiral}
K_{A}^{++}={-\frac{\Nf T_F}{2\pi}}\,\frac{p_-}{p_+},\qquad
K_{A}^{--}={-\frac{\Nf T_F}{2\pi}}\,\frac{p_+}{p_-},\qquad
K_{A}^{+-}=K_{A}^{-+}=0.
\end{equation}
Transversality of the $A$ two-point function instead requires $K_{A}^{+-}=K_{A}^{-+}={\tfrac{\Nf T_F}{2\pi}}$. By combining the chiral $\lambda$s we avoid this anomaly-related problem.

Since the $\lambda$s have conventional couplings to the vector-multiplet fields, a chiral loop of
$n$ segments scales as  $|\ell|^{-n}$ at large momentum and has an integrable simple pole at each soft point. The UV radial integral converges for $n\ge3$. As a result, the convergent fermion triangle in
section~\ref{sec:brst-induced-Aphiphi} and the fermion box
of~\cite{OhlssonSax:2014jtq} can be evaluated directly in two dimensions.

\paragraph{$\phi$ kernels}
Using~\eqref{eq:brst-dirac-scalar-matrix} and~\eqref{eq:brst-fund-propagator-residues}, we evaluate
the scalar loop to get
\begin{equation}\label{eq:brst-scalar-insertion-trace}
\begin{aligned}
 &S_{\lambda_{\sL}}(\ell)S_{\lambda_{\sR}}(\ell+p)
 (\mathcal M_I)^{\dot\alpha}{}_{\beta}
 (\mathcal M_J)^{\beta}{}_{\dot\alpha}
\\
 &\quad+S_{\lambda_{\sR}}(\ell)S_{\lambda_{\sL}}(\ell+p)
 (\mathcal M_I)^{\alpha}{}_{\dot\beta}
 (\mathcal M_J)^{\dot\beta}{}_{\alpha}
 =-\frac{4\delta_{IJ}\,\ell\cdot(\ell+p)}{\ell^2(\ell+p)^2}.
\end{aligned}
\end{equation}
Here the two index structures on $\mathcal M_I$ denote its two diagonal
blocks in~\eqref{eq:brst-dirac-scalar-matrix}.
In DDR we continue the dot product with the same $\widehat D$-dimensional
Dirac-trace prescription used for the gauge kernel. Combining this with the quadratic term
$\tfrac{\Nf}{2}\operatorname{Tr}(S_\lambda V_\phi S_\lambda V_\phi)$
and the scalar metric in footnote~\ref{fn:brst-scalar-metric}, we find
\begin{equation}\label{eq:brst-phi-lambda-start}
 K^{ab;\alpha\dot\alpha}_{\phi\;\beta\dot\beta}(p)
 =-2\Nf T_F\delta^{ab}
 \delta^\alpha{}_\beta\delta^{\dot\alpha}{}_{\dot\beta}
 \mu^{d-\widehat D}\int\!\frac{d^{\widehat D}\ell}{(2\pi)^{\widehat D}}
 \frac{\operatorname{tr}_\gamma[
 \slashed\ell(\slashed\ell+\slashed p)]}
 {\ell^2(\ell+p)^2}\,.
\end{equation}
The gamma-matrix trace gives
\begin{equation}\label{eq:Piphi-trace}
 \operatorname{tr}_\gamma[
 \slashed\ell(\slashed\ell+\slashed p)]
 =2\ell\cdot(\ell+p)\,.
\end{equation}
Combining~\eqref{eq:Piphi-trace} and~\eqref{eq:brst-bubble-dot} gives
\begin{equation}\label{eq:brst-phi-result}
 K^{ab;\alpha\dot\alpha}_{\phi\;\beta\dot\beta}(p)
 =2\Nf T_F\delta^{ab}
 \delta^\alpha{}_\beta\delta^{\dot\alpha}{}_{\dot\beta}
 p^2B_{d,\widehat D}(p).
\end{equation}
The fundamental bosons couple to $\phi$ via
$-\tfrac14H^\dagger\phi_{\alpha\dot\alpha}
\phi^{\alpha\dot\alpha}H$ in the Lorentzian action (the corresponding coefficient in $S_E$ is $+\tfrac14$), so contribute through a tadpole diagram whose integral is $T_1=0$ by \eqref{eq:brst-tadpole-zero}. As a result,~\eqref{eq:brst-phi-result} is the complete scalar kernel.

\paragraph{$\psi$ kernels}
The relevant terms in the Euclidean action involving $\psi_{\sR}$ are $-\sqrt{2}H^\dagger\bar\psi_{\sR}\lambda_{\sL}
+\sqrt{2}\bar\lambda_{\sL}\psi_{\sR}H$~\eqref{eq:brst-euclidean-fields}{.}
The internal propagators are
$G_H(\ell)=2/\ell^2$ and
{$S_{\lambda_{\sL}}(\ell)=-i\ell_-/\ell^2$}~\eqref{eq:brst-fund-propagator-residues}.
The contraction of the epsilon tensors in~\eqref{eq:uv-fund} gives a
minus sign in the kinetic pairing
$\bar\psi_{\sR\,\dot\alpha\dot a}\psi_{\sR}^{\dot\alpha\dot a}$.
Together with the two Yukawa magnitudes $\sqrt2$ and the propagator
residues, this gives the coefficient $+4i$.
For external fields \(\bar\psi_{\sR}(-p)\psi_{\sR}(p)\), the
$H$ line has momentum \(\ell-p\). We change variables $\ell\rightarrow -\ell$ giving the denominator $\ell^2(\ell+p)^2$ and the overall coefficient becomes ${-4i}$.
Using~\eqref{eq:brst-bubble-first}, we obtain
\begin{equation}\label{eq:brst-psi-loop}
\begin{aligned}
 K_{\psi_{\sR}}^{ab}(p)
 &={-4}i\Nf T_F\delta^{ab}
 \mu^{d-\widehat D}\int\!\frac{d^{\widehat D}\ell}{(2\pi)^{\widehat D}}
 \frac{\ell_-}{\ell^2(\ell+p)^2}
 ={+2}i\Nf T_F\delta^{ab}p_-B_{d,\widehat D}(p),
\\
 K_{\psi_{\sL}}^{ab}(p)
 &={-4}i\Nf T_F\delta^{ab}
 \mu^{d-\widehat D}\int\!\frac{d^{\widehat D}\ell}{(2\pi)^{\widehat D}}
 \frac{\ell_+}{\ell^2(\ell+p)^2}
 ={+2}i\Nf T_F\delta^{ab}p_+B_{d,\widehat D}(p)\,.
\end{aligned}
\end{equation}

\paragraph{Inverting the constrained fermion kernel.}
\label{app:fermion-normalization}
The quadratic action~\eqref{eq:brst-vector-quadratic-functional} 
sums over all fermion components, including those related by their {Lorentzian} reality conditions
\begin{equation}\label{eq:fermion-reality}
 (\psi_{\sL}^{\alpha\dot a})^\dagger
 =-\epsilon_{\alpha\beta}\epsilon_{\dot a\dot b}
       \psi_{\sL}^{\beta\dot b}\,,\qquad\qquad
 (\psi_{\sR}^{\dot\alpha\dot a})^\dagger
 =+\epsilon_{\dot\alpha\dot\beta}\epsilon_{\dot a\dot b}
       \psi_{\sR}^{\dot\beta\dot b}\,.
\end{equation}
 To invert it, we have to first express it in terms of
independent components. With the barred fields defined in~\eqref{eq:brst-euclidean-fields}, the continued relations are
\begin{equation}\label{eq:fermion-euclidean-reality}
 \bar\psi_{\sL\,\alpha\dot a}
 =+i\,\epsilon_{\alpha\beta}\epsilon_{\dot a\dot b}
       \psi_{\sL}^{\beta\dot b}\,,\qquad\qquad
 \bar\psi_{\sR\,\dot\alpha\dot a}
 =-i\,\epsilon_{\dot\alpha\dot\beta}\epsilon_{\dot a\dot b}
       \psi_{\sR}^{\dot\beta\dot b}\,,
\end{equation}
rather than ordinary Euclidean complex conjugation. Using $K_\psi(-p)=-K_\psi(p)$, we find
\begin{equation}\label{eq:fermion-full-sum}
 \sum_{\alpha,\dot a}\int \frac{d^dp}{(2\pi)^d}
   \bar\psi_{\sL\,\alpha\dot a}(-p)K_{\psi_{\sL}}(p)
   \psi_{\sL}^{\alpha\dot a}(p)
 =2\sum_{\dot a}\int \frac{d^dp}{(2\pi)^d}
   \bar\psi_{\sL\,1\dot a}(-p)K_{\psi_{\sL}}(p)
   \psi_{\sL}^{1\dot a}(p)\,,
\end{equation}
with an analogous expression for the other chirality.
As a result, we see that the kernel on independent
components is $2K_\psi$. Its inverse is
\begin{equation}\label{eq:fermion-independent-inverse}
 (2K_{\psi_{\sR}})^{-1}_{ab}(p)
 ={-\frac{i\delta^{ab}}{4\Nf T_Fp_-B_{d,\widehat D}(p)}}\,,
 \end{equation}
and with $p_-$ replaced by $p_+$ for the inverse $\psi_L$ kernel. Sending $\widehat D\to d$ and using {$p^2=p_+p_-$} we obtain the fermion
propagators in equation~\eqref{eq:brst-vector-propagators}. 

\paragraph{Auxiliary $D$ kernel}
Writing $D^{\dot a\dot b}=D^r(\sigma^r)^{\dot a\dot b}$, with $\operatorname{tr}(\sigma^r\sigma^s)=2\delta^{rs}$, the $H$ bubble  gives
\begin{equation}\label{eq:brst-D-loop}
\begin{aligned}
 K_D^{ab;rs}(p)
 &=\Nf\left(\frac12\right)^2(2)^2
 \operatorname{tr}(T^aT^b)\operatorname{tr}(\sigma^r\sigma^s)
 \mu^{d-\widehat D}\int\!\frac{d^{\widehat D}\ell}{(2\pi)^{\widehat D}}
 \frac{1}{\ell^2(\ell+p)^2}
\\
 &=2\Nf T_F\delta^{ab}\delta^{rs}
 B_{d,\widehat D}(p)\,.
\end{aligned}
\end{equation}
The factors of $2$ and $1/2$ come from the propagators and vertices respectively, reproducing~\eqref{eq:brst-vector-quadratic-functional}.

\section{The $d$-dimensional $A\phi^2$ vertex}
\label{app:dirac-vertex}
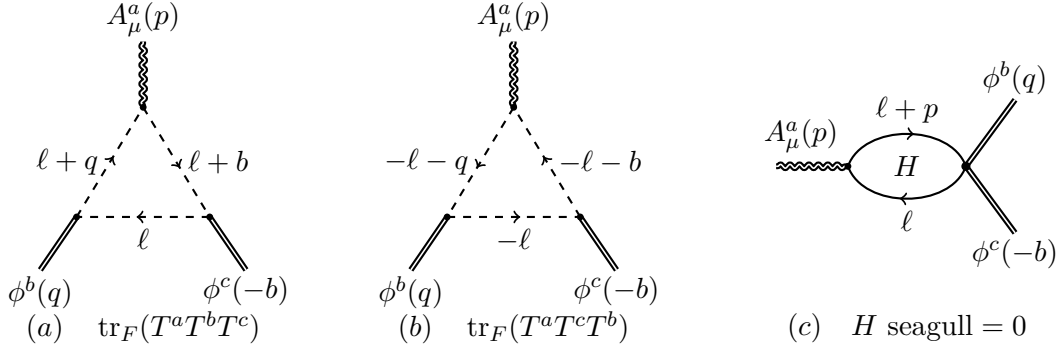
\begin{figure}[htbp]
\centering
\begin{tikzpicture}[
  x=1cm,y=1cm,
  every node/.style={font=\small,inner sep=2pt},
  fund flow/.style={postaction={decorate},
    decoration={markings,mark=at position .55 with {\arrow{>}}}}
]
\begin{scope}[xshift=-5.05cm]
  \coordinate (a) at (0,1.15);
  \coordinate (b) at (-.88,-.30);
  \coordinate (c) at (.88,-.30);
  \draw[gluon] (a) -- (0,2.02);
  \node[above] at (0,2.02) {$A_\mu^a(p)$};
  \draw[scalar] (b) -- (-1.36,-1.00);
  \draw[scalar] (c) -- (1.36,-1.00);
  \node[below] at (-1.36,-1.00) {$\phi^b(q)$};
  \node[below] at (1.36,-1.00) {$\phi^c(-b)$};
  \draw[hyp fermion,fund flow] (b) -- (a)
    node[midway,left=2pt] {$\ell+q$};
  \draw[hyp fermion,fund flow] (a) -- (c)
    node[midway,right=2pt] {$\ell+b$};
  \draw[hyp fermion,fund flow] (c) -- (b)
    node[midway,below=2pt] {$\ell$};
  \foreach \v in {a,b,c} \fill (\v) circle[radius=.045cm];
  \node at (0,-1.76) {$(a)\quad\operatorname{tr}_F(T^aT^bT^c)$};
\end{scope}
\begin{scope}[xshift=-.15cm]
  \coordinate (a) at (0,1.15);
  \coordinate (b) at (-.88,-.30);
  \coordinate (c) at (.88,-.30);
  \draw[gluon] (a) -- (0,2.02);
  \node[above] at (0,2.02) {$A_\mu^a(p)$};
  \draw[scalar] (b) -- (-1.36,-1.00);
  \draw[scalar] (c) -- (1.36,-1.00);
  \node[below] at (-1.36,-1.00) {$\phi^b(q)$};
  \node[below] at (1.36,-1.00) {$\phi^c(-b)$};
  \draw[hyp fermion,fund flow] (a) -- (b)
    node[midway,left=2pt] {$-\ell-q$};
  \draw[hyp fermion,fund flow] (c) -- (a)
    node[midway,right=2pt] {$-\ell-b$};
  \draw[hyp fermion,fund flow] (b) -- (c)
    node[midway,below=2pt] {$-\ell$};
  \foreach \v in {a,b,c} \fill (\v) circle[radius=.045cm];
  \node at (0,-1.76) {$(b)\quad\operatorname{tr}_F(T^aT^cT^b)$};
\end{scope}
\begin{scope}[xshift=5.05cm]
  \coordinate (a) at (-.78,.37);
  \coordinate (s) at (.78,.37);
  \draw[gluon] (-1.73,.37) -- (a);
  \node[above=3pt] at (-1.40,.37) {$A_\mu^a(p)$};
  \draw[scalar] (s) -- (1.43,1.25);
  \draw[scalar] (s) -- (1.43,-.51);
  \node[above] at (1.43,1.25) {$\phi^b(q)$};
  \node[below] at (1.43,-.51) {$\phi^c(-b)$};
  \draw[hyp scalar,fund flow] (a) to[out=70,in=110]
    node[midway,above=2pt] {$\ell+p$} (s);
  \draw[hyp scalar,fund flow] (s) to[out=250,in=290]
    node[midway,below=2pt] {$\ell$} (a);
  \node at (0,.37) {$H$};
  \fill (a) circle[radius=.045cm];
  \fill (s) circle[radius=.060cm];
  \node at (0,-1.76) {$(c)\quad H\text{ seagull}=0$};
\end{scope}
\end{tikzpicture}
\caption{Fundamental graphs for the gauge--scalar--scalar vertex.
All external momenta are incoming, with \(b=p+q\); arrows orient the
labelled internal momenta. The two fermion colour orderings have
opposite numerators, giving
\(\operatorname{tr}_F(T^a[T^b,T^c])\).
The scalar seagull graph vanishes under \(\ell\mapsto-\ell-p\).}
\label{fig:dirac-fundamental-vertex}
\end{figure}

In this appendix we calculate the $A\phi^2$ vertex of the induced theory~\eqref{eq:brst-Aphiphi-d}. We combine fundamental chiral
loops before continuing the numerator algebra in the usual way. The longitudinal part
reduces to an bubble integral while the transverse part
requires a triangle integral.  The contributing graphs are shown in Figure~\ref{fig:dirac-fundamental-vertex}.

\paragraph{Conventions and the fundamental loop.}
We write the defining-loop measure as
\begin{equation}
\int_\ell \equiv \mu^{d-\widehat D}
\int\frac{d^{\widehat D}\ell}{(2\pi)^{\widehat D}}\,,
\end{equation}
and take $\widehat D\to d=2+2\omega$ after integration.
The incoming momenta are $p,q,-b=-p-q$. The scalar legs are the independent real fields $\phi_I^b$ and
$\phi_J^c$; we suppress their common $\delta_{IJ}$ below.
The equivalent bispinor convention is given
in~\eqref{eq:brst-Aphiphi-scalar-metric}.

Writing the propagator and coupling to $A$ for the combined Dirac fermion gives
\begin{equation}\label{eq:dirac-rules}
 S_\lambda(\ell)=-i\,\frac{\slashed\ell}{\ell^2},
 \qquad
 V_A^{a\mu}=-i\gamma^\mu T^a \,,
\end{equation}
The column $\Lambda$ and barred row $\bar\Lambda$ are defined
in~\eqref{eq:brst-euclidean-dirac-operator}. The scalar vertices are
$V_{\phi_I}^a=\mathcal M_I T^a$, obtained from the bispinor
vertex~\eqref{eq:brst-bispinor-scalar-vertex} by the real-component
change of basis given there. The cubic term in the
fermion determinant is
\begin{equation}\label{eq:brst-cubic-dirac-determinant}
 \Gamma_\lambda^{(3)}=-\frac{\Nf}{3}
 \operatorname{Tr}\!\left[(S_\lambda V)^3\right],\qquad
 V=V_A^{a\mu}A^a_\mu+V_{\phi_I}^a\phi_I^a.
\end{equation}
This trace includes colour, the Dirac factor and the internal doublet;
$\operatorname{tr}_\gamma$ below denotes the two-component Dirac trace.
For the $A\phi^2$ term, the three cyclic placements of the gauge vertex
cancel the factor $1/3$.
The numerical propagator and vertex factors give
$(-i)^3(-i)=1$.
The internal-doublet trace reduces to $2\delta_{IJ}$, leaving the
two-component Dirac trace in the numerator below.
The two orderings of the triangle combine into
\(\operatorname{tr}_F(T^a[T^b,T^c])=iT_Ff^{abc}\).
The Euclidean effective-action, with its momentum delta
function removed, can be written as
\begin{equation}\label{eq:dirac-numerator}
 \Gamma^{a\mu}_{bc}=-2i\Nf T_Ff^{abc}I^\mu\,,
\end{equation}
where
\begin{equation}
 I^\mu=\int_\ell
 \frac{N^\mu(\ell)}{\ell^2(\ell+q)^2(\ell+b)^2}\,,
 \qquad \qquad
 N^\mu(\ell)=\operatorname{tr}_\gamma
 [(\slashed\ell+\slashed b)\gamma^\mu
  (\slashed\ell+\slashed q)\slashed\ell]\,.
\end{equation}
Using the four-gamma trace in~\eqref{eq:PiA-Ftrace} gives
\begin{equation}\label{eq:dirac-numerator-expanded}
 N^\mu(\ell)=2\Bigl[
 \ell^\mu(\ell^2-q\cdot b)
 +b^\mu\,\ell\cdot(\ell+q)
 +q^\mu\,\ell\cdot(\ell+b)\Bigr].
\end{equation}
The H-loop graph in Figure~\ref{fig:dirac-fundamental-vertex} is proportional to
\begin{equation}\label{eq:dirac-H-zero}
 \int_\ell\frac{(2\ell+p)^\mu}{\ell^2(\ell+p)^2}=0,
\end{equation}
because the numerator changes sign under $\ell\mapsto-\ell-p$.

\paragraph{The longitudinal part.}
Contracting~\eqref{eq:dirac-numerator-expanded} with $p$ we find
\begin{equation}\label{eq:dirac-longitudinal}
\begin{aligned}
 p_\mu I^\mu
 &=\int_\ell
 \frac{\ell^2[(\ell+b)^2-(\ell+q)^2]
       -q^2(\ell+b)^2+b^2(\ell+q)^2}
      {\ell^2(\ell+q)^2(\ell+b)^2}\\
 &=\int_\ell\left[
 \frac1{(\ell+q)^2}-\frac1{(\ell+b)^2}
 -\frac{q^2}{\ell^2(\ell+q)^2}
 +\frac{b^2}{\ell^2(\ell+b)^2}\right]\\
 &=b^2B_{d,\widehat D}(b)-q^2B_{d,\widehat D}(q)\,.
\end{aligned}
\end{equation}
The first two terms cancel after a shift and the remaining result follows from~\eqref{eq:brst-BdDhat}.

\paragraph{The transverse part.} The transverse part cannot be expressed in terms of $B_{d,\widehat D}$. We combine the propagators using Feynman parameters
\begin{equation}\label{eq:dirac-feynman}
 \frac1{\ell^2(\ell+q)^2(\ell+b)^2}
 =2\int_{\mathcal S}
 \frac1{[x_1\ell^2+x_2(\ell+q)^2+x_3(\ell+b)^2]^3}\,,
\end{equation}
where
\begin{equation}\label{eq:dirac-parameters}
 \int_{\mathcal S}
 \equiv\int_0^1 dx_1\int_0^{1-x_1}dx_2,
 \qquad x_3=1-x_1-x_2 .
\end{equation}
Completing the square in the denominator gives
\begin{equation}\label{eq:dirac-square}
 x_1\ell^2+x_2(\ell+q)^2+x_3(\ell+b)^2
 =(\ell+x_2q+x_3b)^2+\Delta\,,
\end{equation}
where $\Delta=q^2x_1x_2+p^2x_2x_3+b^2x_1x_3$. Shifting the loop momentum to $L=\ell+x_2q+x_3b$ gives
\begin{equation}\label{eq:dirac-shifted-integral}
 I^\mu=2\mu^{d-\widehat D}
 \int_0^1dx_1\int_0^{1-x_1}dx_2
 \int\!\frac{d^{\widehat D}L}{(2\pi)^{\widehat D}}\,
 \frac{N^\mu(L-x_2q-x_3b)}{(L^2+\Delta)^3}\,.
\end{equation}
The numerator can be split into the parts parallel and perpendicular to $p$ by writing
\begin{equation}\label{eq:dirac-decomposition}
 I^\mu=\frac{p^\mu}{p^2}(p\cdot I)+q_\perp^\mu F_T
 \,,\qquad \qquad \mbox{where }\qquad
 q_\perp^\mu=q^\mu-\frac{p\cdot q}{p^2}p^\mu\,.
\end{equation}
Projecting onto the perpendicular part and noting that $q_\perp\cdot q=q_\perp\cdot b=q_\perp^2$ we get
\begin{equation}\label{eq:dirac-projected-numerator}
 \frac{q_{\perp\mu}N^\mu(\ell)}{2q_\perp^2}
 =\frac{q_\perp\cdot\ell}{q_\perp^2}
   (\ell^2-q\cdot b)+2\ell^2+\ell\cdot(q+b).
\end{equation}
Using $\ell=L-x_2q-x_3b$ and the definition of $\Delta$, we obtain
\begin{equation}\label{eq:dirac-shift-relations}
 \frac{q_\perp\cdot\ell}{q_\perp^2}
 =\frac{q_\perp\cdot L}{q_\perp^2}-(1-x_1)\,,\qquad\qquad
 (x_2q+x_3b)^2=x_2q^2+x_3b^2-\Delta .
\end{equation}
Since the denominator depends only on $L^2$, odd powers of $L$ in the numerator integrate to zero, and using rotational symmetry we can replace
$L_\mu L_\nu$ by $\delta_{\mu\nu}L^2/\widehat D$ under the
integral. In particular, the quadratic cross term from the first
product in \eqref{eq:dirac-projected-numerator} simplifies 
\begin{equation}\label{eq:dirac-angular}
 \frac{(q_\perp\cdot L)\,[L\cdot(x_2q+x_3b)]}{q_\perp^2}
 \ \longmapsto\ \frac{1-x_1}{\widehat D}L^2\,.
\end{equation}
Putting all this together we obtain
\begin{equation}\label{eq:dirac-transverse-integral}
 F_T=4\mu^{d-\widehat D}\int_{\mathcal S}
 \int\!\frac{d^{\widehat D}L}{(2\pi)^{\widehat D}}\,
 \frac{\left[1-\frac2{\widehat D}
 +(1+\frac2{\widehat D})x_1\right]L^2
       -x_1\Delta-p^2x_2x_3}{(L^2+\Delta)^3}\,.
\end{equation}
The two scalar loop integrals can now be performed 
\begin{equation}\label{eq:dirac-gaussians}
\begin{aligned}
 \mu^{d-\widehat D}\int\!\frac{d^{\widehat D}L}{(2\pi)^{\widehat D}}
 \frac1{(L^2+\Delta)^3}
 &=\frac{\mu^{d-\widehat D}}{(4\pi)^{\widehat D/2}}
   \frac{\Gamma(3-\widehat D/2)}2\,
   \Delta^{\widehat D/2-3}\,,\\
 \mu^{d-\widehat D}\int\!\frac{d^{\widehat D}L}{(2\pi)^{\widehat D}}
 \frac{L^2}{(L^2+\Delta)^3}
 &=\frac{\mu^{d-\widehat D}}{(4\pi)^{\widehat D/2}}
   \frac{\widehat D\,\Gamma(2-\widehat D/2)}4\,
   \Delta^{\widehat D/2-2}\,.
\end{aligned}
\end{equation}
giving
\begin{equation}\label{eq:dirac-transverse-parameters}
 F_T=
 \frac{\mu^{d-\widehat D}\Gamma(2-\widehat D/2)}
      {(4\pi)^{\widehat D/2}}
 \int_{\mathcal S}\Bigl\{
 [\widehat D-2+(2\widehat D-2)x_1]\Delta^{\widehat D/2-2}
 +(\widehat D-4)p^2x_2x_3\Delta^{\widehat D/2-3}
 \Bigr\}\,.
\end{equation}
The terms proportional to $x_1$ and $p^2x_2x_3$ can be combined using 
\begin{equation}\label{eq:dirac-ibp}
\begin{aligned}
 &\partial_{x_1}\!\left[x_1(1-x_1)\Delta^{\widehat D/2-2}\right]
 -\partial_{x_2}\!\left[x_1x_2\Delta^{\widehat D/2-2}\right]\\
 &\quad=
 \left[\tfrac{\widehat D}{2}-1-(\widehat D-1)x_1\right]
 \Delta^{\widehat D/2-2}
 -\left(\tfrac{\widehat D}{2}-2\right)
 p^2x_2x_3\Delta^{\widehat D/2-3}\,.
\end{aligned}
\end{equation}
The boundary terms cancel in the $x_i$ integrals and so integrating by parts we find
\begin{equation}\label{eq:dirac-ibp-integrated}
 \int_{\mathcal S}\left[
 (2\widehat D-2)x_1\Delta^{\widehat D/2-2}
 +(\widehat D-4)
 p^2x_2x_3\Delta^{\widehat D/2-3}\right]\\
 =(\widehat D-2)
 \int_{\mathcal S}\Delta^{\widehat D/2-2}\,.
\end{equation}
Inserting this relation in~\eqref{eq:dirac-transverse-parameters} we finally arrive at
\begin{equation}\label{eq:dirac-shift}
 F_T=
 \frac{2(\widehat D-2)\mu^{d-\widehat D}\Gamma(2-\widehat D/2)}
 {(4\pi)^{\widehat D/2}}
 \int_{\mathcal S}\Delta^{\widehat D/2-2}
 =8\pi(\widehat D-2)C_{\widehat D+2},
\end{equation}
where
\begin{equation}\label{eq:dirac-triangle}
 C_{\widehat D+2}(p^2,q^2,b^2)=
 \frac{\mu^{d-\widehat D}\Gamma(2-\widehat D/2)}
 {(4\pi)^{\widehat D/2+1}}
 \int_{\mathcal S}
 (q^2x_1x_2+p^2x_2x_3+b^2x_1x_3)^{\widehat D/2-2}\,.
\end{equation}
Combining~\eqref{eq:dirac-longitudinal} and~\eqref{eq:dirac-shift}, gives the vertex
\eqref{eq:brst-Aphiphi-d}.

\end{document}